\documentclass[12pt,a4paper]{article}

\usepackage{amsmath}
\usepackage{amssymb}
\usepackage{hyperref}
\usepackage{hyperref}
\usepackage{graphics}
\usepackage{graphicx}
\usepackage{subfigure}
\usepackage{epsfig}

\let\oldappendix=\appendix
\let\oldsection=\section
\renewcommand{\appendix}{\oldappendix%
\def\theequation{\Alph{section}.\arabic{equation}}%
\renewcommand{\section}{\setcounter{equation}{0}\oldsection}}

\newcommand{\beq}{\begin{equation}}
\newcommand{\eeq}{\end{equation}}
\newcommand{\beqa}{\begin{eqnarray}}
\newcommand{\eeqa}{\end{eqnarray}}
\newcommand{\no}{\nonumber}

\newcommand{\mnod}{\stackrel{\circ}{M}} 
\newcommand{\sfrac}[2]{{\textstyle\frac{#1}{#2}}}

\begin{document}

\hfill 

\hfill 

\bigskip\bigskip

\begin{center}

{{\Large\bf  $\mbox{\boldmath$\eta$}$,  $\mbox{\boldmath$\eta'$}$ 
photo- and \\electroproduction off nucleons}}

\end{center}

\vspace{.4in}

\begin{center}
{\large B. Borasoy\footnote{email: borasoy@physik.tu-muenchen.de},
 E. Marco\footnote{email: emarco@physik.tu-muenchen.de},
 and S. Wetzel}

\bigskip

\bigskip

\href{http://www.ph.tum.de/}{Physik Department}\\
\href{http://www.tum.de/}{Technische Universit{\"a}t M{\"u}nchen}\\
D-85747 Garching, Germany \\

\vspace{.2in}

\end{center}

\vspace{.7in}

\thispagestyle{empty} 

\begin{abstract}
The photo- and electroproduction of the $\eta, \eta'$ mesons on nucleons
are investigated within a relativistic chiral unitary
approach based on coupled channels. The $s$-wave potentials
for electroproduction and meson-baryon scattering are derived from a chiral
effective Lagrangian which includes the $\eta'$ as an explicit degree of
freedom and incorporates important features of the underlying QCD
Lagrangian such as the axial $U(1)$ anomaly. The effective potentials
are iterated in a Bethe-Salpeter equation and cross sections for $\eta,
\eta'$ photo- and electroproduction from nucleons are obtained.
The results for the $\eta'$ photoproduction cross section
on protons reproduce the appearance of an $S_{11}$ resonance
around 1.9 GeV observed at ELSA.
The inclusion of electromagnetic form factors increases 
the predicted $\eta$ electroproduction cross sections on the proton, 
providing a qualitative explanation
for the hard form factor of the photocoupling amplitude observed
at CLAS.

\end{abstract}\bigskip

\begin{center}
\begin{tabular}{ll}
\textbf{PACS:}&11.80.-m, 12.39.Fe, 13.60.-r, 13.75.-n, 14.20.Gk\\[6pt]
\textbf{Keywords:}& $\eta, \eta'$ electroproduction, chiral symmetry,
 unitarity, resonances.
\end{tabular}
\end{center}

% 11.80.-m   Relativistic scattering theory  
% 12.39.Fe   Chiral Lagrangians
% 13.60.-r   Photon and charged-lepton interactions with hadrons
% 13.75.-n   Hadron-induced low- and intermediate-energy reactions 
%            and scattering (energy less than or equal to 10 GeV)
% 14.20.Gk   Baryon resonances with S=0
% 

\vfill

%%%%%%%%%%%%%%%%%%%%%%%%%%%%%%%%%%%%%%%%%%%%%%%%%%%%%%%%%%%%%%%%%%%%%%%%%%%%%%
%%%%%%%%%%%%%%%%%%%%%%%%%%%%%%%%%%%%%%%%%%%%%%%%%%%%%%%%%%%%%%%%%%%%%%%%%%%%%%
\section{Introduction}\label{sec:intro}

Chiral symmetry is believed to govern interactions among hadrons at
low energies where the relevant degrees of freedom are not the quark and 
gluon fields of the QCD Lagrangian, but composite hadrons. In order to
make contact with experiment one must resort to methods
such as chiral perturbation theory (ChPT) which incorporates the symmetries and
symmetry breaking patterns of underlying QCD and is written in terms of the 
active degrees of freedom.
A systematic loop expansion can be carried out which inherently involves a
characteristic scale $\Lambda_\chi = 4 \pi F_\pi \approx 1.2$ GeV at which the
chiral series is expected to break down. The limitation to very low-energy
processes is even enhanced in the vicinity of resonances. The appearance
of resonances in certain channels constitutes a major problem to the
loopwise expansion of ChPT since their contribution cannot be reproduced
at any given order of the chiral series. This can be prevented by including
the resonance exchanges explicitly with the couplings fixed from
electromagnetic and hadronic data. In that case, however, the theory
loses its predictive power and does not provide a stringent test of chiral
symmetry.

Recently, considerable effort has been undertaken to combine the effective
chiral Lagrangian approach with non-perturbative methods, both
in the meson-baryon sector \cite{KSW1,KSW2} and in the purely mesonic
sector \cite{OO}.
The combination with non-perturbative
schemes have made it possible to go to energies beyond $\Lambda_\chi$
and to generate resonances dynamically. Two prominent examples
in the baryonic sector are the $\Lambda(1405)$ and the $S_{11}(1535)$.
The first one is an $s$-wave resonance just below the $K^- p$
threshold and dominates the interaction of the $\bar{K} N$ system.
The properties of the  $\Lambda(1405)$ which appears at the correct
position with the right width were reproduced remarkably well in
\cite{KSW1,OR,OM}. 
A similar analysis was performed in \cite{Inoue} where
in addition to the $S_{11}(1535)$ the $\Delta(1620)$
was also obtained via the inclusion of the $\pi \pi N$ channel.
The $S_{11}(1535)$ is of particular interest since
it decays very strongly into the $\eta N$ channel and provides
insight into the $\eta N$ interaction. Applying the same formalism as in
\cite{KSW1}, the authors in \cite{KSW2} could generate the $S_{11}(1535)$,
which they identify as a quasi-bound $K \Lambda$-$K \Sigma$ state.
Information on the $S_{11}(1535)$ was experimentally extracted from 
precise eta photoproduction data off protons close to threshold at MAMI (Mainz)
\cite{mami} and together with an analogous electroproduction experiment
performed at ELSA (Bonn) \cite{elsa} the data covered the whole range of the
$S_{11}(1535)$ resonance.

In general, photoproduction of mesons is a tool to study baryonic
resonances and the investigation of transitions between these
states provides a crucial test for hadron models. The dominance
of the $\Delta (1232)$ in the photoproduction of
pions, e.g., has allowed to extract information on its electromagnetic
transition amplitudes. 
Because of their hadronic decay modes nucleon resonances have large overlapping
widths, which makes it difficult to study individual states, but selection
rules in certain decay channels can reduce the number of possible resonances.
The isoscalars $\eta$ and $\eta'$ are such examples since, due to isospin
conservation, only the isospin-$\frac{1}{2}$ excited states decay into the
$\eta N$ and $\eta' N$ channels.

Electroproduction experiments are even more sensitive to the structure
of the nucleon due to the longitudinal coupling of the virtual photon to the
nucleon spin and might in addition yield some insight into the possible
onset of perturbative QCD. Perturbative QCD should apply at sufficiently
high photon virtuality $Q^2=-k^2$, see e.g.\ \cite{CP, LDW}, however there
is no consensus about how high the momentum transfer must be. It has been
found experimentally, that in the case of
electroproduction of the $\Delta (1232)$ resonance at momentum transfers up to
$Q^2 = 4.0$ GeV$^2$ perturbative QCD is not applicable, \cite{eldel}, 
whereas a possible onset of scaling in the reaction 
$ e +p \rightarrow e+p + \eta$ at $Q^2 = 3.6$ GeV$^2$ is reported in
\cite{els11}. 
CLAS at JLab has also electroproduced $\eta$ mesons for invariant
momentum transfers $Q^2$ between 0.375 and 1.375 GeV$^2$ \cite{thomp}.
It was found that the $S_{11}$ photocoupling 
$A_{1/2}$ decreases much slower  with $Q^2$ than, e.g., the nucleon
dipole form factor, which is unusual and difficult to explain theoretically.

Furthermore, there is still some controversy about the nature of
the $\eta NN$ and $\eta' NN$ couplings.
For $\eta$ photoproduction in the $S_{11} (1535)$ resonance region, e.g.,   
both an effective Lagrangian approach \cite{BeM} and
coupled channel models \cite{Be,Ti} have been employed.
In these approaches the coupling of the $\eta$ to the nucleons is described by
both a pseudovector and a pseudoscalar term and the coupling constant and the
coupling structure of the Born terms is unknown. In \cite{Ti} it has been 
shown that differential cross sections are rather sensitive to the
assumptions about this  vertex, but within the framework
of chiral perturbation theory this coupling is fixed at lowest
order by making use of the chiral $SU(3)_L \times SU(3)_R$
symmetry of the Lagrangian, whereas explicitly chiral
symmetry breaking terms appear at higher orders. The $SU(3)_L \times SU(3)_R$
symmetric limit provides therefore a convenient starting point
which overcomes the problem of fixing the $\eta NN$ vertex.
The $SU(3)$ chiral meson-baryon Lagrangian has been used in a coupled channel
model \cite{KWW} and by adjusting a few parameters a large amount of 
low-energy data was described. All the above mentioned investigations
have in common that they treat the
$\eta$ meson as a pure $SU(3)$ octet state $\eta_8$ and mixing of $\eta_8$ with
the corresponding singlet state  $\eta_0$ which yields the physical states
$\eta$ and  $\eta'$ is neglected.

The  $\eta'$ is interesting by itself. The QCD Lagrangian with massless quarks
exhibits an $SU(3)_L \times SU(3)_R$ chiral symmetry which is broken down
spontaneously to $SU(3)_V$, giving rise to a Goldstone boson octet of
pseudoscalar mesons which become massless in the chiral limit of zero
quark masses. On the other hand, the axial $U(1)$ symmetry of the QCD
Lagrangian is broken by the anomaly. 
The corresponding pseudoscalar singlet would otherwise have a mass comparable
to the pion mass \cite{W}. Such a particle is missing in the spectrum and the
lightest candidate would be the $\eta'$ with a mass of 958 MeV which is
considerably heavier than the octet states.
In conventional chiral perturbation theory the $\eta'$ 
is not included explicitly, although it does show up in the form of a
contribution to an LEC.
However, it is also possible to include the $\eta'$ explicitly
in the chiral Lagrangian framework as has been done, e.g.,
in \cite{leut, H-S} for the purely
mesonic sector which was then extended to include baryons in \cite{B}.
Within this framework the $\eta'$ is combined with the 
Goldstone bosons $(\pi, K, \eta)$ into a nonet and the $\eta NN$ and
$\eta' NN$ couplings are constrained by chiral symmetry.

The $\eta'$ photoproduction has been investigated theoretically in
\cite{ZMB, Li, BWW}. In the effective Lagrangian approach of
\cite{ZMB} a pseudoscalar coupling of the $\eta'$ to the nucleons
was chosen and it was concluded that the $\eta' N$ decay channel is
dominated by the not so well established $D_{13}(2080)$
resonance, whereas in the quark model used in \cite{Li} the off-shell 
effects of the $S_{11} (1535)$ were prominent. 
In contrast, the experimental data from ELSA
\cite{Pl} suggested the coherent excitation of two resonances 
$S_{11} (1897)$ and $P_{11} (1986)$. In \cite{BWW} $\eta$ and $\eta'$
photoproduction has been studied in the coupled channel formalism working 
along the lines of \cite{KWW}. However, the treatment of
the $\eta'$ is incomplete and only the leading terms in
the meson-baryon potentials are taken into account.
Their results fail to reproduce the $\eta$ photoproduction data
and are unable to describe the appearance
of a $S_{11}$ resonance for $\eta'$ photoproduction as reported at ELSA.
In order to predict the cross sections
for electroproduction,  which will become
available from experiments at Jefferson Lab in the near future \cite{N},
a much more thorough investigation is needed.

In the present work we will focus on $\eta, \eta'$ photo- and
electroproduction off nucleons in a framework which allows a unifying
description of these processes with chiral symmetry and unitarity being the
main ingredients.
This is achieved by combining the effective Lagrangian with a
non-perturbative scheme based on coupled channels and the
Bethe-Salpeter equation.
The approach contains only a few parameters, both coupling constants in the
Lagrangian, so-called low-energy constants (LECs),
and finite range parameters
which appear in the evaluation of the loop integrals. With this small
set of parameters it will be a highly non-trivial task to reproduce
the data on meson photo- and electroproduction
and meson-baryon scattering experiments
which are described simultaneously within this multi-channel analysis.
Hence this investigation will provide a test whether processes up to
energies of $\sqrt{s} \sim 2$ GeV
are still constrained by chiral symmetry 
and whether the $\eta'$ meson can be included in the
effective Lagrangian with baryons as proposed in \cite{B}.

We will perform a global fit to a wide range of meson-baryon scattering
and photoproduction data. To this end,
we restrict ourselves to $s$-waves and therefore the
comparison with data should only be valid in the near threshold region.
At higher energies $p$-waves start dominating, as can be seen, e.g.,
in the work by Caro Ramon et al.\ \cite{KWW}.
One of the purposes of this work is to shed some light on the
$s$-wave resonance $S_{11} (1897)$ which can be studied in the $s$-wave
approximation.
Our results must be compared to the cross section reported in \cite{Pl}.
We will continue giving predictions for $\eta$ and $\eta'$
electroproduction processes and compare our results with
the $\eta$ electroproduction data from CLAS \cite{els11}.
As we will see, the present framework also provides an at least qualitative
explanation of the slow fall-off with increasing $Q^2$ of the
$S_{11} (1535)$ photocoupling. Effects of $\eta$-$\eta'$ mixing
and the importance of the $\eta'$ contributions
in this coupled channel formalism even for processes where it appears
only as a virtual state are discussed in detail.
This may eventually lead to a better understanding of gluonic effects and
the significance of the axial $U(1)$ anomaly in low-energy hadron physics.

In the next two sections, we introduce the effective Lagrangian and the
coupled channel formalism which is then generalized to electroproduction
processes. The results of our analysis are presented and compared to existing
$\eta, \eta'$ photoproduction and $\eta$ electroproduction data in Sec.\ 5,
where also predictions for $\eta'$ electroproduction 
are made. We summarize our findings in Sec.\ 6.

%%%%%%%%%%%%%%%%%%%%%%%%%%%%%%%%%%%%%%%%%%%%%%%%%%%%%%%%%%%%%%%%%%%%%%%%%%%%%%%%

\section{The effective $\mbox{\boldmath$U(3)$}$ Lagrangian}  \label{sec:Lagran}

In this section, we will shortly review a systematic way of including
the $\eta'$ in the chiral effective Lagrangian. For details the reader
is referred to \cite{KL,BB} in the purely mesonic sector or \cite{B}
the presence of baryons.

The $U(3)_L \times U(3)_R$ chiral effective Lagrangian of
the pseudoscalar meson nonet $(\pi,K,\eta_8,\eta_0)$ 
coupled to the ground state
baryon octet $(N,\Lambda, \Sigma, \Xi)$ can be decomposed as
\begin{equation}
{\cal L} = {\cal L}_\phi + {\cal L}_{\phi B}
\end{equation}
with the mesonic piece up to  second chiral order \cite{BB}
\begin{equation}  \label{mes}
{\cal L}_\phi = 
- \frac{v_0}{f_\pi^2} \eta_0^2 + \frac{f_\pi^2}{4}  \langle u_{\mu}
u^{\mu} \rangle + \frac{f_\pi^2}{4} \langle \chi_+ \rangle + 
   i \frac{v_3}{f_\pi} \eta_0 \langle \chi_- \rangle
\end{equation}
and a part ${\cal L}_{\phi B}$ which describes the meson-baryon
interactions \cite{B}. At lowest order it reads
\begin{eqnarray}  \label{bar}
{\cal L}_{\phi B}^{(1)} &=& i \langle \bar{B} \gamma_{\mu} [D^{\mu},B] \rangle 
 - \mnod \langle \bar{B}B \rangle + i u_1 \frac{\eta_0^2}{f_\pi^2}
 \Big(  \langle [D^{\mu},\bar{B}]  \gamma_{\mu}  B \rangle - 
 \langle \bar{B}  \gamma_{\mu}  [D^{\mu},B] \rangle \Big) \no \\
&&  - \frac{1}{2} D \langle \bar{B} \gamma_{\mu}
 \gamma_5 \{u^{\mu},B\} \rangle  
- \frac{1}{2} F \langle \bar{B} \gamma_{\mu} \gamma_5 [u^{\mu},B] \rangle 
- \frac{1}{2} D_s  \langle \bar{B} \gamma_{\mu} \gamma_5 B \rangle 
  \langle u^{\mu} \rangle ,
\end{eqnarray}
where only the terms that are necessary for the present
calculation  are kept
and $\langle \ldots \rangle$ denotes the trace in flavor space.
The pseudoscalar meson nonet is summarized in
$u_{\mu} = i u^\dagger \nabla_{\mu} U u^\dagger$
with 
\begin{equation}
 U(\varphi,\eta_0) = u^2 (\varphi,\eta_0) = 
\exp \left( \sqrt{2} i \frac{\varphi}{f_\pi} + i \sqrt{\frac{2}{3}} \frac{\eta_0}{f_\pi}  \right)
\end{equation}
where $f_\pi \simeq 92.4$ MeV is the pion decay constant and $\varphi$ contains
the Goldstone bosons $(\pi,K,\eta_8)$.
The covariant derivative of the meson fields includes the coupling to an
external photon field 
\begin{equation}
\nabla_{\mu} U = \partial_\mu U - i v_\mu U + i U v_\mu = \partial_\mu U 
+ i e {\cal A}_\mu [Q, U]
\end{equation} 
with $Q = \frac{1}{3} \mbox{diag}(2,-1,-1)$
being the quark charge matrix and ${\cal A}_\mu$ the photon field.
Explicit chiral symmetry breaking is induced via the quark mass matrix
${\cal M} = \mbox{diag}(m_u, m_d,m_s)$
which enters in the combinations $\chi_\pm = 2 B_0 (u^\dagger {\cal M}
u^\dagger \pm u {\cal M} u  )$
with $B_0 = - \langle  0 | \bar{q} q | 0\rangle/ f_\pi^2$ the order
parameter of the spontaneous symmetry violation.

The second and third term of Eq.\ (\ref{mes}) appear already
in conventional ChPT whereas the first and fourth one are due
to the axial $U(1)$ anomaly. The first one is the mass term of the
singlet field $\eta_0$ which remains in the chiral limit of vanishing
quark masses. The coefficient $v_0 $ is a parameter
not fixed by chiral symmetry and in the large $N_c$ limit it is
proportional to the topological susceptibility of Gluodynamics. The
fourth term yields $\eta_8$-$\eta_0$
mixing, which can be described in terms of a single mixing angle,
see Eq.\ (\ref{App:mix}).\footnote{However, as was shown in 
\cite{BB}, it is not the only contribution
to $\eta_8$-$\eta_0$ since terms from the fourth order Lagrangian
yield off-diagonal elements for the derivative pieces of the Lagrangian
so that the mixing in the $\eta$-$\eta'$ system can not be
parameterized in terms of a single mixing angle if large $N_c$
counting rules are not imposed. The presentation of counterterms
of the fourth order mesonic Lagrangian is beyond the scope of this
work and for our purposes it will be sufficient to assume a
mixing scheme in terms of one mixing parameter.}

The ground state baryon octet $(N,\Lambda,\Sigma,\Xi)$ is summarized in a 
$3 \times 3$ matrix $B$, $\mnod$ is the common baryon octet mass in the
chiral limit and $D,F, D_s$ are the axial vector couplings of the baryons
to the mesons. The values of $D$ and $F$ are extracted phenomenologically
from the semileptonic hyperon decays and a fit to data delivers $D= 0.80 
\pm 0.01$, $F=0.46 \pm 0.01$ \cite{CR}. The parameters $u_1$ and $D_s$
do not enter in conventional ChPT and, since
their values have not yet been
determined, we will extract them from our fit.
Finally, the covariant derivative of the baryon fields is given by
\begin{equation}
[ D_\mu, B] = \partial_\mu B + [ \Gamma_\mu, B] 
\end{equation}
with the chiral connection
\begin{equation} \label{gama}
\Gamma_\mu = \sfrac{1}{2} [ u^\dagger,  \partial_\mu u]  + i e 
{\cal A}_\mu  Q .
\end{equation}
Note that there is no pseudoscalar coupling of $\eta_0$ to the baryons of
the form $\eta_0 \bar{B}  \gamma_5 B$. Such a term is in principle possible but
can be absorbed by the $D_s$-term in Eq. (\ref{bar}) 
by means of the equation of motion for the baryons.

At next-to-leading order the terms relevant for $s$-wave meson-baryon
scattering are
\begin{eqnarray}  \label{bar2}
{\cal L}_{\phi B}^{(2)} &=&  b_D \langle \bar{B}  \{\chi_+,B\} \rangle +
b_F \langle \bar{B}  [\chi_+,B] \rangle + b_0 \langle \bar{B}B \rangle
\langle \chi_+ \rangle \no \\
&&
+ i \frac{c_D}{f_\pi} \eta_0 \langle \bar{B}  \{\chi_-,B\} \rangle 
+ i \frac{c_F}{f_\pi} \eta_0 \langle \bar{B}  [\chi_-,B] \rangle 
+ i \frac{c_0}{f_\pi} \eta_0 \langle \bar{B} B\rangle \langle \chi_+
\rangle \no \\
&&
+ d_1 \langle \bar{B} \{u_{\mu},[u^{\mu},B]\}  \rangle 
+ d_2 \langle \bar{B} [u_{\mu},[u^{\mu},B]]  \rangle
+ d_3 \langle \bar{B} u_{\mu} \rangle  \langle u^{\mu} B  \rangle
+ d_4 \langle \bar{B} B \rangle  \langle u^{\mu}  u_{\mu} \rangle \no \\
&&
+ d_5 \langle \bar{B} \{ u_{\mu}, B \}  \rangle  \langle u^{\mu} \rangle
+ d_6 \langle \bar{B} [ u_{\mu}, B ]  \rangle  \langle u^{\mu} \rangle
+ d_7 \langle \bar{B} B \rangle  \langle u^{\mu} \rangle \langle 
u_{\mu} \rangle .
\end{eqnarray}
We made use of the Cayley-Hamilton identity, in order to eliminate 
$\langle \bar{B} \{u_{\mu},\{u^{\mu},B\}\}  \rangle $.
The terms $b_D,b_F, b_0$ and $d_{1, \ldots,4}$ are already present in
the $SU(3)$ case,
whereas  $c_D,c_F, c_0$ and $d_{5,6,7}$ are new terms of the extended theory.
The LECs $b_D$ and $b_F$ are responsible for the splitting of the
baryon octet masses 
at leading order in symmetry breaking. Working in the $SU(2)$ limit of equal
$u$ and $d$ quark masses, $m_u = m_d = \hat{m}$, one obtains
\begin{eqnarray} \label{massdiff}
M_\Sigma - M_N &=& 4 (b_D -b_F) (m_K^2 -m_\pi^2) \no \\
M_\Xi - M_N &=& - 8 b_F (m_K^2 -m_\pi^2) \no \\
M_\Sigma - M_\Lambda &=& \frac{16}{3} b_D (m_K^2 -m_\pi^2).
\end{eqnarray}
Since the three baryon mass differences are represented in terms of
two parameters,
there is a corresponding sum rule -- the Gell-Mann--Okubo mass relation
for the baryon octet \cite{GMO}:
\begin{equation}
M_\Sigma - M_N = \frac{1}{2} (M_\Xi - M_N) + \frac{3}{4}
(M_\Sigma - M_\Lambda )
\end{equation}
which experimentally has only a 3\% deviation.
A least-squares fit to
the mass differences (\ref{massdiff}) yields
$b_D = 0.066 $ GeV$^{-1}$ and $b_F = -0.213 $ GeV$^{-1}$.
The $b_0$ term, on the other hand, cannot be determined from the masses alone.
One needs further information which is provided by the pion-nucleon
$\sigma$-term and reads at leading order
\begin{equation}
\sigma_{\pi N} = \hat{m} \langle N |   \bar{u} u + \bar{d} d | N \rangle 
=  - 2 m_\pi^2 ( b_D +b_F + 2 b_0).
\end{equation}
Employing the empirical value of \cite{GLS}, $\sigma_{\pi N}= 45 \pm 8$ MeV,
one obtains $b_0= -0.52 \pm 0.10$ GeV$^{-1}$. Recently, this value has
been questioned \cite{PSWA} and the authors of this work arrive at a
value  $\sigma_{\pi N}= 60 \pm 7$ MeV which would translate into a
value of $b_0= -0.71 \pm 0.09 $ GeV$^{-1}$. In particular the
latter value yields a large strangeness content of the proton, however
both values may change if loop effects are included \cite{BM}. 
Due to these uncertainties in the value of $b_0$, any
fitted value which lies in the range $-0.80$ 
GeV$^{-1}< b_0 < -0.20$ GeV$^{-1} $ is still acceptable.
In the $U(3)$ formalism three more explicitly symmetry breaking terms enter, 
$c_D, c_F$ and $c_0$, which have unknown values. These are suppressed by
one order of 1/$N_c$ with respect to their counterparts $b_D, b_F$ and $b_0$,
and following 1/$N_c$ arguments one can assume them to be significantly smaller
in magnitude, e.g.\ $|c_0| \ll |b_0|$. We will vary these three parameters only
within  small ranges around zero and compare the results with existing data.

The situation is less clear for the derivative terms $d_1 , \ldots, d_7$.
In \cite{KWW} the parameters $d_1$ to $ d_4$ were determined in
a coupled channel analysis
while being subject to two constraints from $\pi N$ and $K N$
scattering lengths.
The authors obtain the values $d_1 = -0.20$, $d_2 = 0.22$, $d_3 = 0.42$ and
$d_4 = -1.62$ in units of GeV$^{-1}$. 
However, they employ the $SU(3)$ chiral Lagrangian which does not include the 
$\eta'$ explicitly and work in the heavy baryon formulation treating
the baryons
as heavy sources while loop diagrams are regularized by an explicit cutoff.
We expect the values of $d_1, \ldots, d_4$ to change in our approach since we
derive the potentials from the relativistic $U(3)$ Lagrangian and 
the loops are evaluated in dimensional regularization.
An estimate for the parameters $d_{1,\ldots,7}$ can alternatively be obtained
by assuming resonance saturation. Within this model
the decuplet $T$ of spin-$\frac{3}{2}$ baryon fields which
includes the $\Delta(1232)$ is expected to determine approximately
the values of the $d_{1,\ldots,7}$ parameters.
The interaction of the decuplet states with the octet baryons and
the pseudoscalar
mesons reads at lowest order
\begin{equation}
{\cal L}_{T B \phi} = \frac{{\cal C}}{2} \bar{T}_\mu u^\mu B + \; h.c. 
\end{equation}
where the coupling constant $|{\cal C}|=$ 1.2--1.8
can be determined from
the decays $T \rightarrow B \pi$. 
The decuplet contributions to the LECs $d_{1,\ldots,7}$ yield 
at leading order in the averaged decuplet mass $M_\Delta$ via
resonance saturation
the values $d_1 = -0.27$, $d_2 = 0.09$, $d_3 = 0.27$, $d_4 = -0.27$,
$d_5 = 0$, $d_6 = 0.18$ and $d_7 = 0.09$ in units of GeV$^{-1}$ for
$|{\cal C}|= 1.5$ and $M_\Delta = 1.38$ GeV.
While $d_{1,2,3}$ are in qualitative agreement with the fitted
values from \cite{KWW}, the values for $d_4$ differ considerably.
Contributions from other resonances have been neglected here since we are only
interested in a rough order of magnitude estimate for these parameters.
We will therefore vary  $d_i$, $i=1,2,3,5,6,7$, within small
ranges around the values
Most unknown couplings are constrained within certain
ranges and they will be fixed from a fit to many scattering data.
After setting up the Lagrangian which we will be using in this work, we can now
proceed by explaining our coupled channel approach.

%%%%%%%%%%%%%%%%%%%%%%%%%%%%%%%%%%%%%%%%%%%%%%%%%%%%%%%%%%%%%%%%%%%%%%%%%%%%%%%%%%

\section{The coupled channel approach}

The Lagrangian of the preceding section could in principle be
used to calculate the one-loop diagrams of $\eta'$ electroproduction,
but the perturbative expansion of the chiral series will be useless
for this case. First, the $\eta'$ with a mass of 958 MeV close to the
scale of chiral symmetry breaking $\Lambda_\chi \approx 1.2$ GeV 
introduces a new massive scale in the effective theory which spoils
the strict chiral counting scheme unless one imposes large $N_c$
counting rules \cite{KL}\footnote{For specific processes such
as the dominant hadronic decay mode of the $\eta'$,
$\eta' \rightarrow  \eta \pi \pi$, it has been shown that the usage of 
infrared regularization yields a well-behaved chiral series
since loops with an $\eta'$ are suppressed \cite{BB2}}.
Large $N_c$ counting rules in turn imply that the $\eta'$ mass must
be treated as a small quantity. In our opinion, this assumption is
phenomenologically not justified and we consider the $\eta'$ to
be a massive state. Second, $\eta'$ electroproduction has a
threshold close to $\sqrt{s} \approx 2$ GeV far beyond the scale
$\Lambda_\chi \approx 1.2$ GeV at which the perturbative
chiral expansion is expected to break down. 
One must therefore resort to non-perturbative schemes.

\begin{figure}[ht]
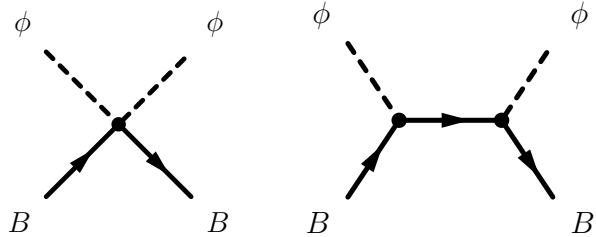

\centering 
\includegraphics[width=3.cm]{fig1a.eps} \qquad
\includegraphics[width=3.9cm]{fig1b.eps}
\caption{Contact interaction and Born direct term
         for meson-baryon scattering.
         Solid and dashed lines denote the baryons and pseudoscalar mesons,
         respectively.}
\label{figmes}
\end{figure}

For the investigation of hadronic resonances the combination of the effective
chiral Lagrangian with coupled channel approaches have been proven
to be useful. E.g., in \cite{KSW2} the $S_{11}(1535)$ nucleon resonance
emerged as a quasi-bound state of $K \Lambda$ and $K \Sigma$.
In the present investigation, we employ a relativistic chiral unitary approach 
based on coupled channels. 
By imposing constraints from unitarity we perform the
resummation of the amplitudes obtained from the
tree level potentials and the loop integrals.
Let us describe first the coupled channel approach for meson-baryon scattering,
which will then be generalized to electroproduction of mesons.
To this end, one expands $u_\mu$ and $\Gamma_\mu$ in Eqs. (\ref{mes}),
(\ref{gama}) in terms of the mesons
\begin{equation}
 u_\mu = - \frac{2}{f_\pi} \partial_\mu \varphi - \frac{1}{f_\pi}
\sqrt{\frac{2}{3}}  \partial_\mu \eta_0 + \ldots  \; ,
\end{equation}
\begin{equation}
 \Gamma_\mu = \frac{1}{2 f_\pi^2} [ \varphi,
 \partial_\mu \varphi]+ \ldots
\end{equation}
In our model the relativistic tree level amplitude $V_\alpha^\beta$ for
the meson-baryon scattering process $B_a \phi_i \rightarrow B_b \phi_j$
is obtained by the diagrams
shown in Fig.\ (\ref{figmes}).

In the present work, we restrict ourselves to the $s$-wave partial
wave amplitude $V$ which is given by
\begin{equation}
V(s) = \frac{1}{ 8 \pi} \sum_{\sigma =1}^2 \int d \Omega \; T(s,\Omega;\sigma)
\end{equation}
where we have averaged over the spin $\sigma$ of the baryons and $s$ is
the invariant energy squared.
It is most convenient to work in the isospin basis and to characterize the
meson-baryon states in the electroproduction process 
$\gamma N \rightarrow \phi B$ by their total isospin $I=1/2$ or  $I=3/2$.
The analysis reduces then to five channels with total isospin $I= 1/2$ which
are $|\pi N \rangle^{(1/2)}$, $|\eta N \rangle^{(1/2)}$, 
$|K \Lambda \rangle^{(1/2)}$, $|K \Sigma \rangle^{(1/2)}$, 
$|\eta' N \rangle^{(1/2)}$ (labeled with indices 1, 2, 3, 4, 5, 
respectively) and two channels with $I=3/2$, 
$|\pi N \rangle^{(3/2)}$, $|K \Sigma \rangle^{(3/2)}$ 
(labeled by indices 6 and 7). The potentials read
\begin{equation}  \label{ampl}
V_\alpha^\beta = \frac{N_a N_b}{f_\pi^2} C_\alpha^\beta
\end{equation}
where $\alpha$ and $\beta$ label the channels involved and
$N_{a(b)} = \sqrt{E_{a(b)} + M_{a(b)}}$ and $E_{a(b)}$,
$M_{a(b)}$ are the energy and mass of the incoming (outgoing) baryon.
Note that we have replaced the common
baryon octet mass $\mnod$ by the physical masses $M_{a(b)}$ which is consistent
at the order the potentials $V$ are calculated
and use of the physical masses is also mandatory in order
to reproduce the correct threshold positions of the different
channels involved.
The coefficients $C_{\alpha}^{\beta}(s)$  are given in App.\ \ref{app:pot}.

For each partial wave $l$ unitarity imposes a restriction on
the (inverse) $T$-matrix above the pertinent thresholds
\begin{equation} \label{unit}
\mbox{Im} T^{-1}_l = - \frac{|\mbox{\boldmath$q$}_{cm}|}{8 \pi \sqrt{s}} 
\end{equation}
with
$\mbox{\boldmath$q$}_{cm}$ being the three-momentum in the
center-of-mass frame of the channel under consideration.
Hence the imaginary part of $T^{-1}$ is similar to the imaginary piece
of the fundamental scalar loop integral $\tilde{G}$
above threshold
\begin{equation}
\tilde{G}(q^2) = \int \frac{d^d l}{(2 \pi)^d} 
\frac{i}{[ (q-l)^2 - M_B^2 + i \epsilon]
   [ l^2 - m_\phi^2 + i \epsilon] } 
\end{equation}
with $M_B$ and $m_\phi$ being the physical masses of
the baryon and the meson, respectively.
For the finite part, $G$ of $\tilde{G}$,
one obtains, e.g., in dimensional regularization
\begin{eqnarray}
G(q^2) &=& \frac{1}{32 \pi^2 q^2} \Bigg\{ q^2
\left[ \ln\Big(\frac{m_\phi^2 }{\mu^2}\Big) +
\ln\Big(\frac{M_B^2 }{\mu^2}\Big) -2 \right] 
+ (m_\phi^2 - M_B^2)  \ln\left(\frac{m_\phi^2 }{M_B^2}\right) \no \\
&&  - 8 \sqrt{q^2}|\mbox{\boldmath$q$}_{cm}| \; \mbox{artanh }
\left(\frac{2 \sqrt{q^2}|\mbox{\boldmath$q$}_{\scriptstyle{cm}}|}{
(m_\phi + M_B )^2 - q^2} \right) \Bigg\}
\end{eqnarray}
where $\mu$ is the regularization scale. We allow this scale
to vary for the different loops, in order to simulate higher
order contributions.
Alternatively, in \cite{OM,Inoue} the real piece has been adjusted by
introducing a scale dependent constant for each channel in analogy
to a subtraction constant of a dispersion relation for $T^{-1}$ . 
In a more general way, one could model the real parts by taking any analytic
function in $s$ and the baryon and meson masses.
This option has been successfully applied for the case of $SU(2)$
ChPT in \cite{NA}, but a straightforward generalization to the
$U(3)$ case would introduce a number of unknown new parameters
and thus reduce the predictive power of the present approach.

The inverse of the amplitude $T^{-1}$ can be decomposed into real
and imaginary parts
\begin{equation} \label{invers}
T^{-1} = \tau^{-1} + G
\end{equation}
where $\tau$ and Re$[G]$ give the real part and Im$[G]$ gives
the imaginary part required by unitarity, Eq.\ (\ref{unit}).
Inverting  (\ref{invers}) yields
\begin{equation} \label{tau}
T = [1 + \tau \cdot G]^{-1} \; \tau
\end{equation}
which is understood to be a matrix equation. The matrix $G$ is diagonal
and includes the expressions for the loop integrals in each channel.
Expanding expression (\ref{tau})
\begin{equation}
T = \tau - \tau \cdot G \cdot \tau \ldots
\end{equation}
and from matching to our tree level amplitude it follows

\begin{equation}
\tau = V .
\end{equation}
Our final expression for the $T$ matrix reads then
\begin{equation}
T = [1 + V \cdot G]^{-1} \; V
\end{equation}
which amounts to a summation of a bubble chain in the $s$-channel.
This is equivalent to a Bethe-Salpeter equation with $V$ as
potential.

%%%%%%%%%%%%%%%%%%%%%%%%%%%%%%%%%%%%%%%%%%%%%%%%%%%%%%%%%%%%%%%%%%%%%%%%%%%%%%%
\section{Extension to photo- and electroproduction}  \label{sec:ext}
We are now in a position to extend the above developed formalism to
electroproduction processes. Following \cite{KWW} we assume that the
$s$-wave electroproduction process can be described
by a similar Bethe-Salpeter equation as for the strong interactions.
Our starting point are the contact, Born and meson-pole diagrams for
meson electroproduction with vertices of the 
Lagrangian as presented in Sec.~\ref{sec:Lagran}, see Fig.~2.
\begin{figure}
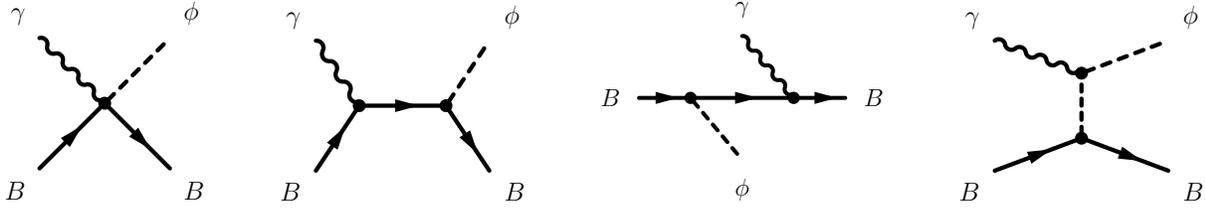

\centering 
\includegraphics[width=2.7cm]{fig2a.eps} \qquad
\includegraphics[width=3.3cm]{fig2b.eps} \qquad
\includegraphics[width=3.8cm]{fig2c.eps} \qquad
\includegraphics[width=3.3cm]{fig2d.eps}
\caption{Contact interaction, Born terms in the $s$- and 
         $u$-channel and meson pole diagram for meson electroproduction.
         Solid and dashed lines denote the baryons and pseudoscalar mesons,
         respectively. The photon is represented by a wavy line.}
\label{figele}
\end{figure}
{}From these diagrams we extract the  transverse and longitudinal
$s$-wave multipoles at leading order which are identified with the 
electroproduction potentials $B_{0+}$ and $C_{0+}$, respectively.
For the proton the expressions  $B_{0+}$ in the isospin basis,
which can be obtained from the formulas appearing
for example in \cite{Belectro}, read
\begin{eqnarray}
B_{0+}^{(1)} &=& (D+F) ( 2 R^{(1)} - S^{(1)} ) \no \\
B_{0+}^{(2)} &=& -\Big(\sqrt{2} ( 2D + 3 D_s) \sin \vartheta
+ (D-3F)\cos \vartheta \Big) S^{(2)} \no \\
B_{0+}^{(3)} &=& (D+3F) R^{(3)} \no \\
B_{0+}^{(4)} &=& (D-F) ( R^{(4)} - 2 S^{(4)} ) \no \\
B_{0+}^{(5)} &=& \Big(\sqrt{2} ( 2D + 3 D_s) \cos \vartheta
- (D-3F)\sin \vartheta \Big) S^{(5)} \no \\
B_{0+}^{(6)} &=&  \sqrt{2} (D+F) ( R^{(6)} + S^{(6)} ) \no \\
B_{0+}^{(7)} &=& -\sqrt{2} (D-F) (  R^{(7)} +  S^{(7)} ) .
\end{eqnarray}
where $\vartheta$ is the mixing angle (see the Appendix).
The functions $R^{(\alpha)}$ and $S^{(\alpha)}$
for the electroproduction process $\gamma B_a \rightarrow \phi B_b$
in the channel $\alpha$ are given by
\begin{eqnarray}
R^{(\alpha)} &=&  \frac{e ( M_a + M_b) N_a}{64 \pi f_\pi \sqrt{3s}
|\mbox{\boldmath$k$}|^2 N_b} \bigg(  \frac{4N_b^2|\mbox{\boldmath$k$}|^2}{M_a
+ \sqrt{s}} + k^2 - 2 E_\phi k_0 \no \\
&&            - \frac{1}{4 |\mbox{\boldmath$k$}| |\mbox{\boldmath$q$}|  }      
          \big[ (k^2-2 E_\phi k_0 )^2 - 4 |\mbox{\boldmath$k$}|^2 
|\mbox{\boldmath$q$}|^2 \big]
          \ln  \frac{2 E_\phi k_0 - 2|\mbox{\boldmath$k$}| 
|\mbox{\boldmath$q$}|-k^2 }{
           2 E_\phi k_0 + 2|\mbox{\boldmath$k$}|
 |\mbox{\boldmath$q$}|-k^2 }  \;  \bigg)
\end{eqnarray}
\begin{eqnarray}
S^{(\alpha)} &=&  \frac{e ( M_a + M_b) N_a}{64 \pi f_\pi \sqrt{3s}
|\mbox{\boldmath$k$}|^2 N_b}
           \bigg( m_\phi^2 + M_a^2 - M_b^2 - 2 E_a E_\phi \no \\
&&        + \frac{2 |\mbox{\boldmath$k$}|^2}{N_a^2 (M_a + \sqrt{s})}
\Big[ (M_a + \sqrt{s})^2 - 2 N_a^2 N_b^2  \Big] \no \\
&&          + \frac{1}{4 |\mbox{\boldmath$k$}| |\mbox{\boldmath$q$}| 
 } \Big((M_a^2-M_b^2- 2 E_a E_\phi + m_\phi^2)^2 
         + 4 |\mbox{\boldmath$k$}|^2  ( 
(\sqrt{s} -M_a)(E_b + M_b) - 
|\mbox{\boldmath$q$}|^2)  \no \\
&&   \qquad \qquad     +\frac{2 |\mbox{\boldmath$k$}|^2 }{N_a^2} 
(\sqrt{s} + M_a) (M_a^2-M_b^2- 2 E_a E_\phi + m_\phi^2)\Big)     \no \\
&& \qquad \qquad \qquad \qquad    \times   \ln \frac{M_b^2 -M_a^2 -m_\phi^2 + 2 E_a E_\phi + 
2|\mbox{\boldmath$k$}| |\mbox{\boldmath$q$}| }{
         M_b^2 -M_a^2 -m_\phi^2 + 2 E_a E_\phi - 2|\mbox{\boldmath$k$}|
 |\mbox{\boldmath$q$}| }   \; \bigg)
\end{eqnarray}
with $\mbox{\boldmath$k$},\mbox{\boldmath$q$} $ the three-momenta of
the photon and the meson in the center-of-mass frame. The energies of
the baryons and meson are given by $E_{a(b)}$ and $E_\phi$, respectively,
whereas $M_{a(b)}$ and $m_\phi$ denote their masses.  The function
$R^{(\alpha)}$ describes charged meson electroproduction processes
on protons, whereas $S^{(\alpha)}$ is obtained when neutral mesons
are produced. In the case of the neutron one obtains
\begin{eqnarray}\label{eq:bn}
B_{0+}^{(1)} &=& -2 (D+F) ( R^{(1)} + S^{(1)} ) \no \\
B_{0+}^{(4)} &=& 2 (D-F) ( R^{(4)} + S^{(4)} ) \no \\
B_{0+}^{(6)} &=& - \sqrt{2} (D+F) ( R^{(6)} + S^{(6)} ) \no \\
B_{0+}^{(7)} &=& \sqrt{2} (D-F) (  R^{(7)} +  S^{(7)} ) \no \\
B_{0+}^{(2)} &=& B_{0+}^{(3)} = B_{0+}^{(5)} = 0 .
\end{eqnarray}
The expressions for the longitudinal $s$-wave potentials
$C_{0+}$ have a similar structure,
but are much more lengthy and will therefore not be presented for brevity.

Once a meson and a baryon have been electroproduced, they may
rescatter into all possible meson-baryon pairs, as shown in
Figs.~3 and ~4. We sum this infinite interaction chain in the
Bethe-Salpeter approach to obtain the final results
for the electric dipole amplitude $E_{0+}$ and the longitudinal
$s$-wave $L_{0+}$. This procedure ensures that the resonances
that appear in meson electroproduction are the same ones
that appear in meson-baryon scattering.
\begin{figure}[ht]
\[
\parbox{2cm}{\centering\includegraphics[scale=0.8]{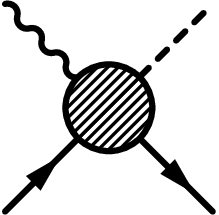}}
=
\parbox{2cm}{\centering\includegraphics[scale=0.8]{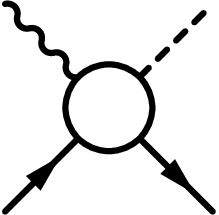}}
+
\parbox{3.6cm}{\centering\includegraphics[scale=0.8]{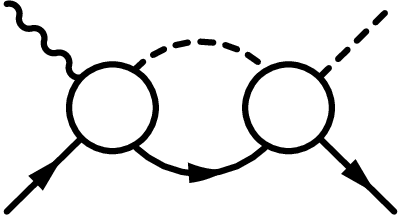}}
+
\parbox{4.6cm}{\centering\includegraphics[scale=0.8]{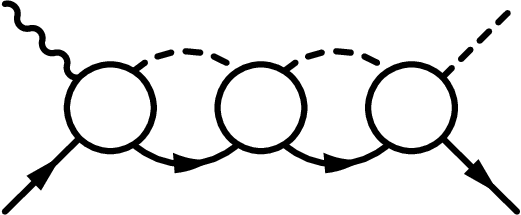}}
+\,\,\ldots
\]
\caption{The $s$-wave electroproduction amplitude is the result
of the tree level electroproduction potentials plus the diagrams
where the final meson-baryon pair are rescattered.
The empty circles denote either the
$s$-wave potentials $B_{0+}$, $C_{0+}$ for electroproduction or
the meson-baryon scattering potentials $V$.}
\end{figure}
\begin{figure}[ht]
\[
\parbox{3cm}{\centering\includegraphics{fig3a.eps}}
=
\parbox{3cm}{\centering\includegraphics{fig3b.eps}}
+
\parbox{5cm}{\centering\includegraphics{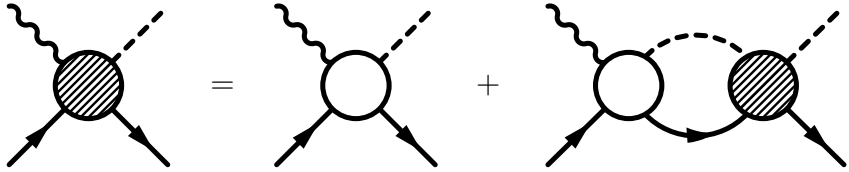}}
\]
\caption{The iterated sum of diagrams in Fig.~3 is illustrated
in compact form.}
%\label{figele}
\end{figure}
The electric dipole moment $E_{0+}$ is given by
\begin{equation}
E_{0+} = [1 + V \cdot G]^{-1} \; B_{0+} \; ,
\end{equation}
while the longitudinal $s$-wave $L_{0+}$ reads
\begin{equation}
L_{0+} = [1 + V \cdot G]^{-1} \; C_{0+} .
\end{equation}
The total cross section for the electroproduction of mesons on the nucleon is
\begin{equation}
\sigma_{tot} = 8 \pi \frac{\sqrt{s} |\mbox{\boldmath$q$}|}{s -M_N^2}
\left(  |E_{0+}|^2  + \epsilon_L |L_{0+}|^2 \right) ,
\end{equation}
with $\epsilon_L = -4 \epsilon s k^2 (s-M_N^2+k^2)^{-2}$ where
$\epsilon$ is the virtual photon polarization.

Within this  formalism the baryons and mesons are treated as
pointlike particles. However, with increasing momentum transfers
the composite structure of hadrons will become important, affecting
the $Q^2$ behavior of the initial electroproduction potentials $B_{0+}$
and $C_{0+}$. In order to analyze these effects, we will
compare in the next section the results for pointlike particles with those
for composite hadrons by inserting monopole
form factors in $B_{0+}$ and $C_{0+}$.

\begin{figure}[t]
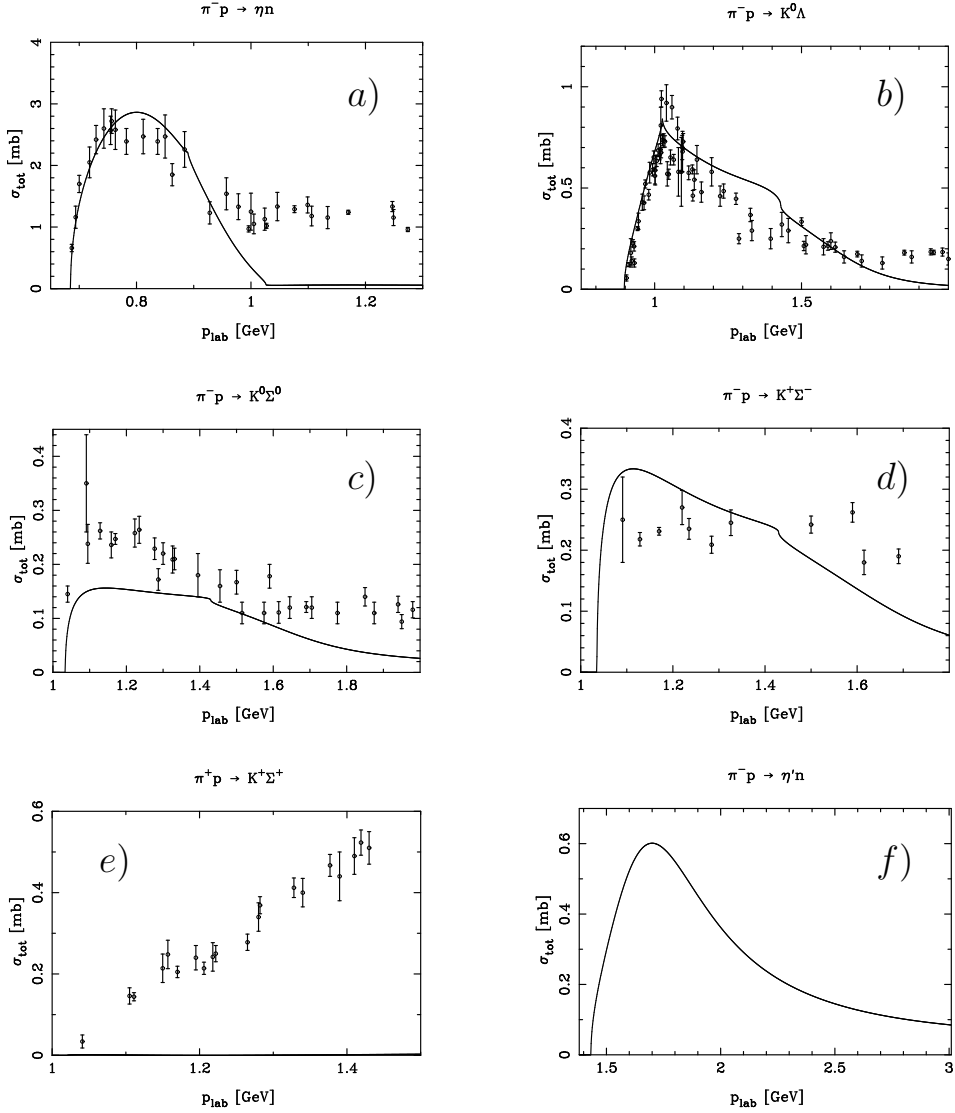

\centering
%\leavemode
\begin{picture}(300,435)
\put(0,310){\makebox(100,120){\epsfig{file=fig5a.ps,width=4.5cm,angle=-90}}}
\put(200,310){\makebox(100,120){\epsfig{file=fig5b.ps,width=4.5cm,angle=-90}}}
\put(0,165){\makebox(100,120){\epsfig{file=fig5c.ps,width=4.5cm,angle=-90}}}
\put(200,165){\makebox(100,120){\epsfig{file=fig5d.ps,width=4.5cm,angle=-90}}}
\put(0,20){\makebox(100,120){\epsfig{file=fig5e.ps,width=4.5cm,angle=-90}}}
\put(200,20){\makebox(100,120){\epsfig{file=fig5f.ps,width=4.5cm,angle=-90}}}
\put(100,395){\large $a)$}
\put(300,395){\large $b)$}
\put(100,250){\large $c)$}
\put(300,250){\large $d)$}
\put(6,105){\large $e)$}
\put(300,105){\large $f)$}
\end{picture}
\caption{Total cross sections for pion-proton collision processes. The data
are taken from \cite{bald}.}
\label{fig:pip}
\end{figure}

%%%%%%%%%%%%%%%%%%%%%%%%%%%%%%%%%%%%%%%%%%%%%%%%%%%%%%%%%%%%%%%%%%%%%%%%%%%%%%%
\section{Results}

In this section we will present the results of our calculation.
We start by performing a global fit to available data for 
meson-proton and photon-proton reactions which constrains the parameters
in the approach. This allows us to give  predictions for further processes
such as the cross sections for $\pi^- p \rightarrow \eta' n$ and
$\eta$ and $\eta'$ electroproduction. We conclude this section by studying
the effects of $\eta$-$\eta'$
mixing and the contributions of the $ \eta'$ meson to those reactions
where it is not produced as a final particle.

\subsection{Fit to the data}

We have performed a global fit to a large amount of data, consisting of
meson-proton and photon-proton reactions for values of $\sqrt{s}$
between 1.5 and 2.0 GeV.
Our cross sections include only $s$-wave
contributions, which are dominant in this energy range
for most processes. From comparison with the work of \cite{KWW} we
expect $p$-waves to exceed the $s$-wave contribution for most
processes at high energies, further away from threshold,
but for a few channels they constitute the main contribution in the
whole energy domain.
This explains why our results are below the experimental
data for some of the discussed processes.

\begin{figure}
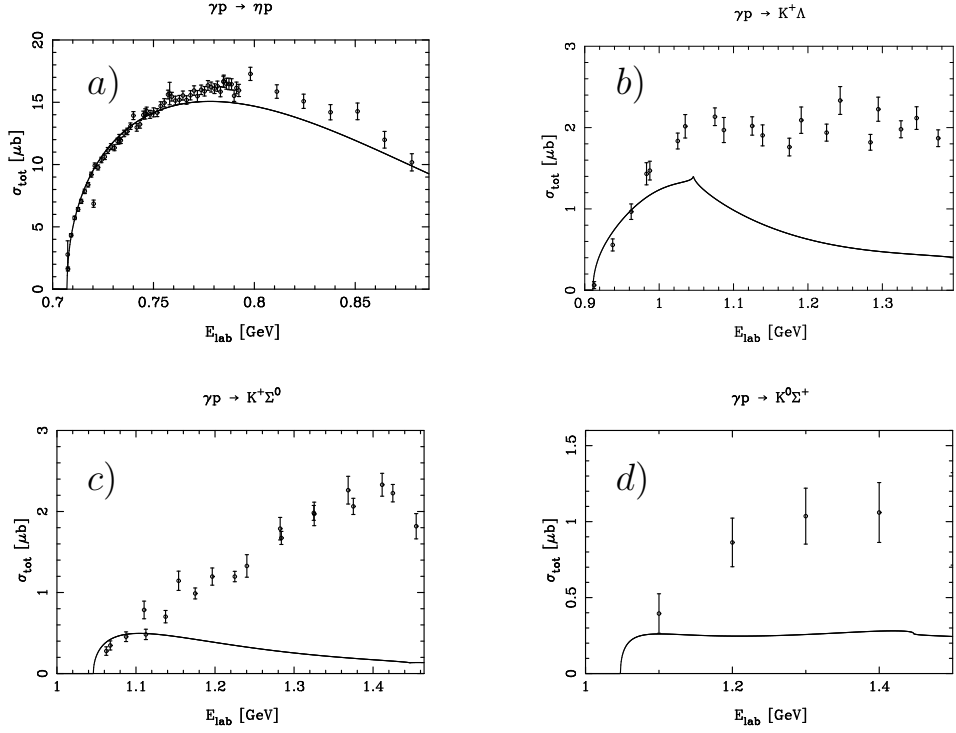

\centering
%\leavemode
\begin{picture}(300,265)
\put(0,165){\makebox(100,120){\epsfig{file=fig6a.ps,width=4.5cm,angle=-90}}}
\put(200,165){\makebox(100,120){\epsfig{file=fig6b.ps,width=4.5cm,angle=-90}}}
\put(0,20){\makebox(100,120){\epsfig{file=fig6c.ps,width=4.5cm,angle=-90}}}
\put(200,20){\makebox(100,120){\epsfig{file=fig6d.ps,width=4.5cm,angle=-90}}}
\put(0,255){\large $a)$}
\put(200,255){\large $b)$}
\put(0,105){\large $c)$}
\put(200,105){\large $d)$}
\end{picture}
\caption{Shown are the total cross sections for $\eta$ and kaon photoproduction
off the proton. The data are taken from
\cite{mami, elsa, tran, goers}.}
\label{fig:gamp}
\end{figure}

The fit yields the values $u_1=-0.0125$ and $D_s=-0.38$ for the
unknown parameters of the leading order Lagrangian, Eq.\ (\ref{bar}), 
$b_D=0.066$ GeV$^{-1}$, $b_F=-0.185$ GeV$^{-1}$
and $b_0=-0.250$ GeV$^{-1}$ for the mass counterterms, and 
$d_1= -0.20$, $d_2=0.10$, $d_3=0.28$, $d_4=-0.25$, $d_5=-0.01$,
$d_6=0.075$ and $d_7=0.075$ (in units of GeV$^{-1}$)
for the derivative couplings  in the next-to-leading order Lagrangian.
For the regularization scales our fitted values are $\mu_{\pi N} = 0.30$ GeV,
$\mu_{\eta N} = 0.77$ GeV, $\mu_{K \Lambda} = 0.14$ GeV, 
$\mu_{K \Sigma} = 1.20$ GeV and $\mu_{\eta' N} = 0.40$ GeV.
The parameters $c_D, c_F$ and $c_0$ are set to zero, since according to large
$N_c$ estimates they are expected to be suppressed and small variations of
their values around zero have not shown any substantial impact on the
cross sections.

The pion-proton cross sections are summarized in Fig.\ \ref{fig:pip}:
Fig.\ \ref{fig:pip}.a  shows
the cross section for the reaction $\pi^- p \rightarrow \eta n$. 
At $p_{\mbox{\scriptsize lab}}$  energies below 1 GeV the
cross section is dominated by the $s$-wave resonance $S_{11}(1535)$,
which is nicely reproduced in our calculation. 
In Fig.\ \ref{fig:pip}.b the results
for the reaction $\pi^- p \rightarrow K^0 \Lambda$ are given. Again the
cross section is dominated by $s$-waves, but in this case
the main contribution stems from the $S_{11}(1650)$ which is accompanied
by a cusp effect
due to the opening of the $K \Sigma$ threshold.
In Figs.\ \ref{fig:pip}.c and \ref{fig:pip}.d the $K \Sigma$
production reactions are depicted. They receive some (small) contribution
from the isospin $T=3/2$ component of the amplitude
and are dominated in the low-energy
region by $s$-wave contributions, but do not signal the appearance
of a resonance. In Fig.\ \ref{fig:pip}.e we
show the cross section for the reaction $\pi^+ p \rightarrow K^+ \Sigma$.
This reaction receives only contributions from the isospin
$T=3/2$ amplitude, and it turns out that the $s$-wave
contribution is negligible as already observed in \cite{KWW}. The last
pion-proton cross section
is shown in Fig.\ \ref{fig:pip}.f, where the $\eta'$ is produced
in $\pi^- p$ collisions. 
Albeit we show this plot with the rest of the pion-proton
cross sections,
it is important to note that
the last reaction is {\it not} included in our fit.
Once the involved parameters are constrained by the other reactions, our model
provides a prediction for this process
which can be checked by future experiments.
\begin{figure}
\centering
\includegraphics[width=8cm,angle=-90]{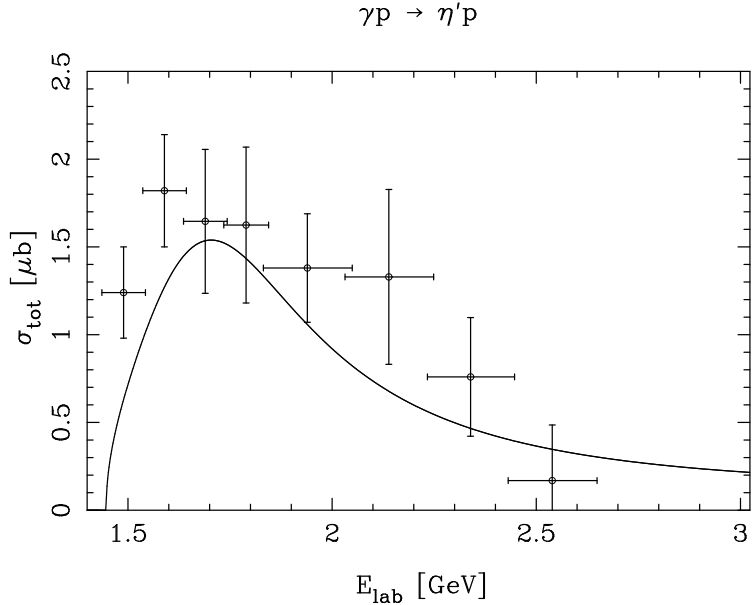}
\caption{Given is the total cross section for $\eta'$-photoproduction
of the proton and the data are taken from \cite{Pl}.}
\label{fig:fig7}
\end{figure}

Let us now turn to the photoproduction cross sections which are shown
in Fig.\ \ref{fig:gamp}. We present in Fig.\ \ref{fig:gamp}.a the $\eta$
photoproduction data measured at MAMI \cite{mami} and ELSA \cite{elsa}
and our fitted result confirms 
the dominance of the $S_{11}(1535)$, which is responsible
for almost the entire cross section  in the low-energy region.
Results for the reaction $\gamma p \rightarrow K^+ \Lambda$ are given
in Fig.\ \ref{fig:gamp}.b, where at low
energies the cross section is dominated by the $S_{11}(1650)$,
while $p$-waves become important for $E_{\mbox{\scriptsize lab}}$
energies above 1.1 GeV. One also observes a cusp effect due to the
$K \Sigma$ threshold.
For the next two reactions, $K^+ \Sigma^0$ and $K^0 \Sigma^+$ photoproduction,
the cross sections are $p$-wave dominated, and our results are able to
account for  small $s$-wave cross sections, describing the data only
in the very near threshold region.
Finally, we show in Fig.\ \ref{fig:fig7} our results for the reaction
$\gamma p \rightarrow \eta' p$, which has been measured at ELSA \cite{Pl},
where the coherent contribution of two resonances, the $S_{11}(1897)$
and $P_{11} (1986)$, was observed.
Our formalism is capable of reproducing
the appearance of an $s$-wave resonance around 1900 GeV in contradistinction
to the work reported in \cite{BWW} which remains far below the measured
decay rates.

So far, we have discussed --with the exception of 
$\pi^- p \rightarrow \eta' n$ in Fig.\ \ref{fig:pip}.f-- only the
channels that were used in order to constrain the undetermined chiral
parameters in our approach. Using the same values for the parameters
we can now predict the cross sections for further processes which will provide
a true test for the applicability of the model. The electroproduction of
$\eta$ and $\eta'$ mesons on a nucleon, e.g., is an ideal testing ground for
the approach, being much more sensitive to its details.

%%%%%%%%%%%%%%%%%%%%%%%%%%%%%%%%%%%%%%%%%%%%%%%%%%%%%%%%%%%%%%%%%%%%%%%%%%%
\subsection{Electroproduction of $\eta$ and $\eta'$ mesons} 
\label{subsec:electro}

\begin{figure}
\centering
%\leavemode
\begin{picture}(330,125)
\put(0,20){\makebox(100,120){\epsfig{file=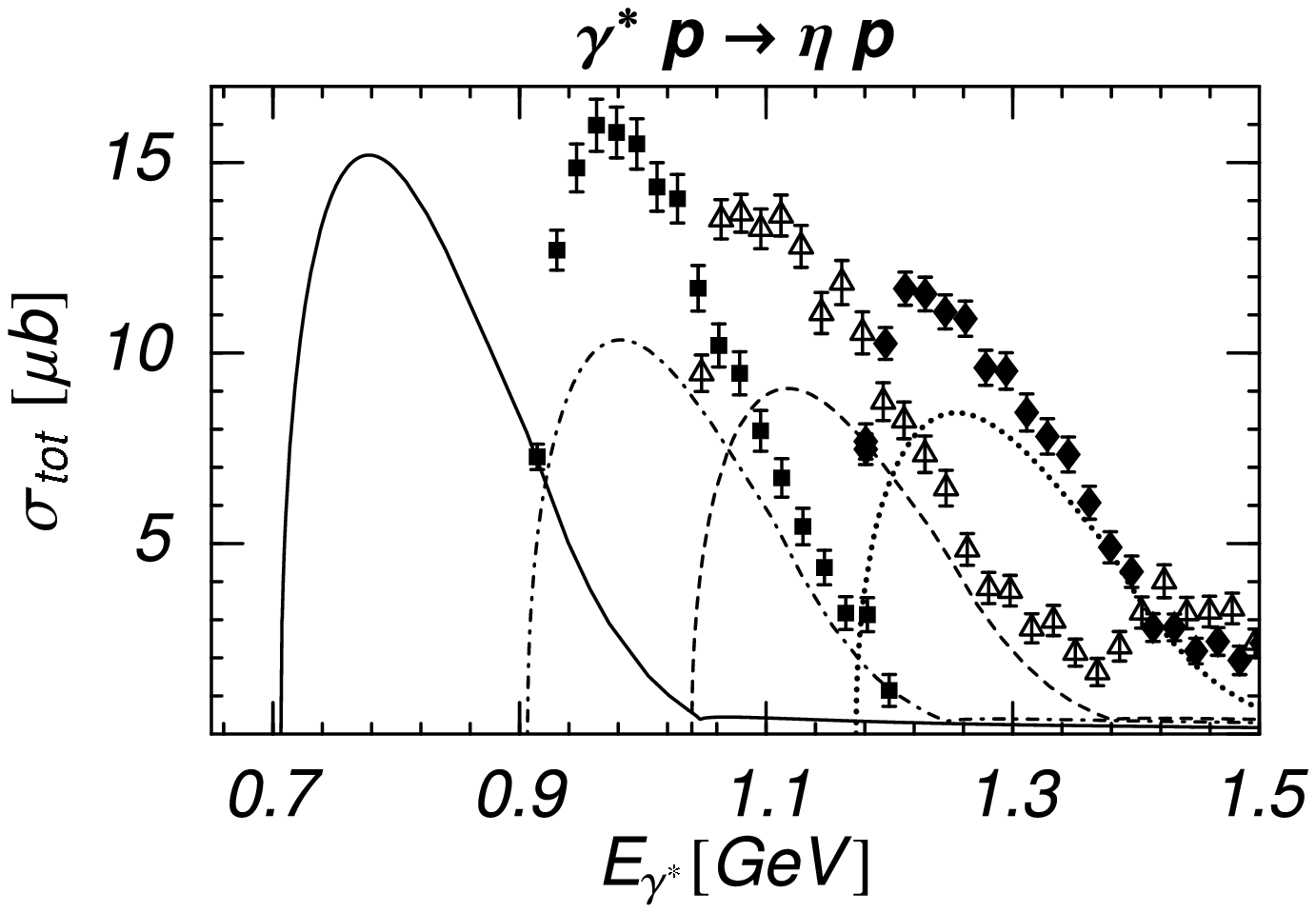,width=7cm}}}
\put(220,20){\makebox(100,120){\epsfig{file=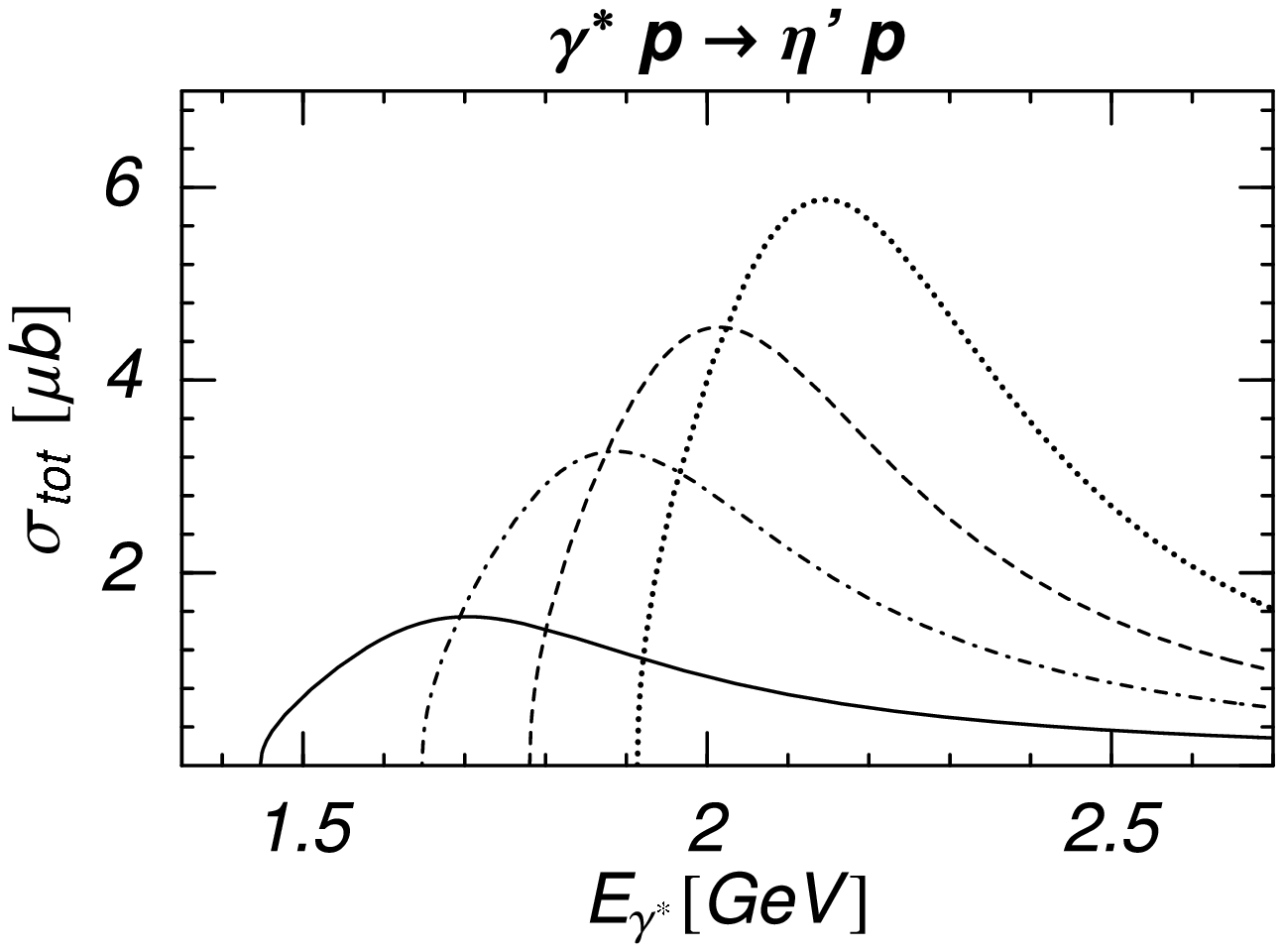,width=7cm}}}
\put(115,115){$a)$}
\put(332,115){$b)$}
\end{picture}
\caption{Cross sections for $\eta$ and $\eta'$ electroproduction on the proton
for various invariant momentum transfers $Q^2$. The different lines refer
to the following values of $Q^2$: solid line: photoproduction ($Q^2=0$);
dash-dotted line and squares: $Q^2=0.375$ GeV$^2$; dashed line and
triangles: $Q^2=0.625$ GeV$^2$; dotted line and diamonds: $Q^2=0.875$ GeV$^2$.
The data for $\eta$ electroproduction with the same $Q^2$ values  
are taken from \cite{thomp}, and only statistical errors are shown.}
\label{fig:elp}
\end{figure}

Detailed information about the electromagnetic structure of the nucleon is
supplied by electroproduction experiments, which yield important constraints
for hadron models. After having fixed the parameters in our approach, we can
now compare our predictions for $\eta$ and $\eta'$ electroproduction with
available experimental data. We first show our predictions
for pointlike hadrons.
The $\eta$ electroproduction on the proton
has been measured in detail at CLAS at JLab \cite{thomp} and the data
is shown together with our results in Fig.\ \ref{fig:elp}.a.

The invariant momentum transfer $Q^2$ of the presented data
ranges from 0.375 to 0.875 GeV$^2$ and the applicability of our approach
to such high momentum transfers may be regarded as questionable.
Nevertheless, we should be able to capture qualitative features of
the $Q^2$ evolution of the cross section, in particular its
slow fall-off which is unusual and in sharp contrast, e.g., to the fall
given by a nucleon dipole form factor. Our results, %on the other hand, 
although showing less decrease with $Q^2$ than that of a simple
nucleon dipole form factor, produce a faster
reduction at low $Q^2$ than the experimental data, and then
flatten at higher momentum transfers.
We employed in all plots a virtual photon polarization of $\epsilon =0.5$
which is roughly the average of the polarizations of the CLAS experiment
and small variations around this value yield changes that are almost
negligible.

Our predictions for electroproduction of the $\eta'$ on the proton
are depicted in  Fig.\ \ref{fig:elp}.b for the same range of momentum
transfers and data for this process will soon become 
available from experiments at Jefferson Lab \cite{N}.
In this case, the features of the $Q^2$ evolution of the cross section
are even more striking, since they exhibit a fast increase instead of
the usual decrease. 

As already mentioned in Sec.~\ref{sec:ext}, the composite structure of
baryons and mesons will become increasingly
important with rising invariant momentum transfers $Q^2$.
It is therefore natural to include form factors in the initial 
electroproduction potentials $B_{0+}$ and $C_{0+}$, in order to account
for the electromagnetic structure of hadrons. One may be inclined
to multiply the potentials by an overall form factor which is the same for all
hadrons that interact with the photon (and thus the same 
for all channels). Obviously, this would reduce all
electroproduction cross sections, e.g., it would be easy to produce
a decreasing $Q^2$ evolution of the $\eta'$ electroproduction
cross section, in accordance with other
electroproduction processes. However, this procedure will also 
yield smaller cross sections for $\eta$ electroproduction 
leading to stronger disagreement with the data.
In this case, the inclusion of form factors which accounts for the
electromagnetic structure of the baryons and mesons seems to worsen
the situation.

As we will see now, there is indeed a way of including form factors for
the particles, while at the same time improving the situation {\it both}
for $\eta$ {\it and} $\eta'$ electroproduction.
Based on the observation that form factors provide a more realistic description
of the electromagnetic response of hadrons, we will present a plausible
qualitative explanation for the unusually hard transition form factor of
the $S_{11}(1535)$. This hard form factor has been difficult to understand
theoretically, and only recent work within a constituent quark model using
a hypercentral potential led to better agreement with experiment \cite{gian}.
At the same time, it has been claimed
that the hard form factor is in contrast to models that interpret the
$S_{11}(1535)$ as a quasi-bound $K \Lambda$-$K \Sigma$ state \cite{burk}.
Here, we will show that this is not necessarily the case.

\begin{figure}
\centering
%\leavemode
\begin{picture}(330,125)
\put(0,20){\makebox(100,120){\epsfig{file=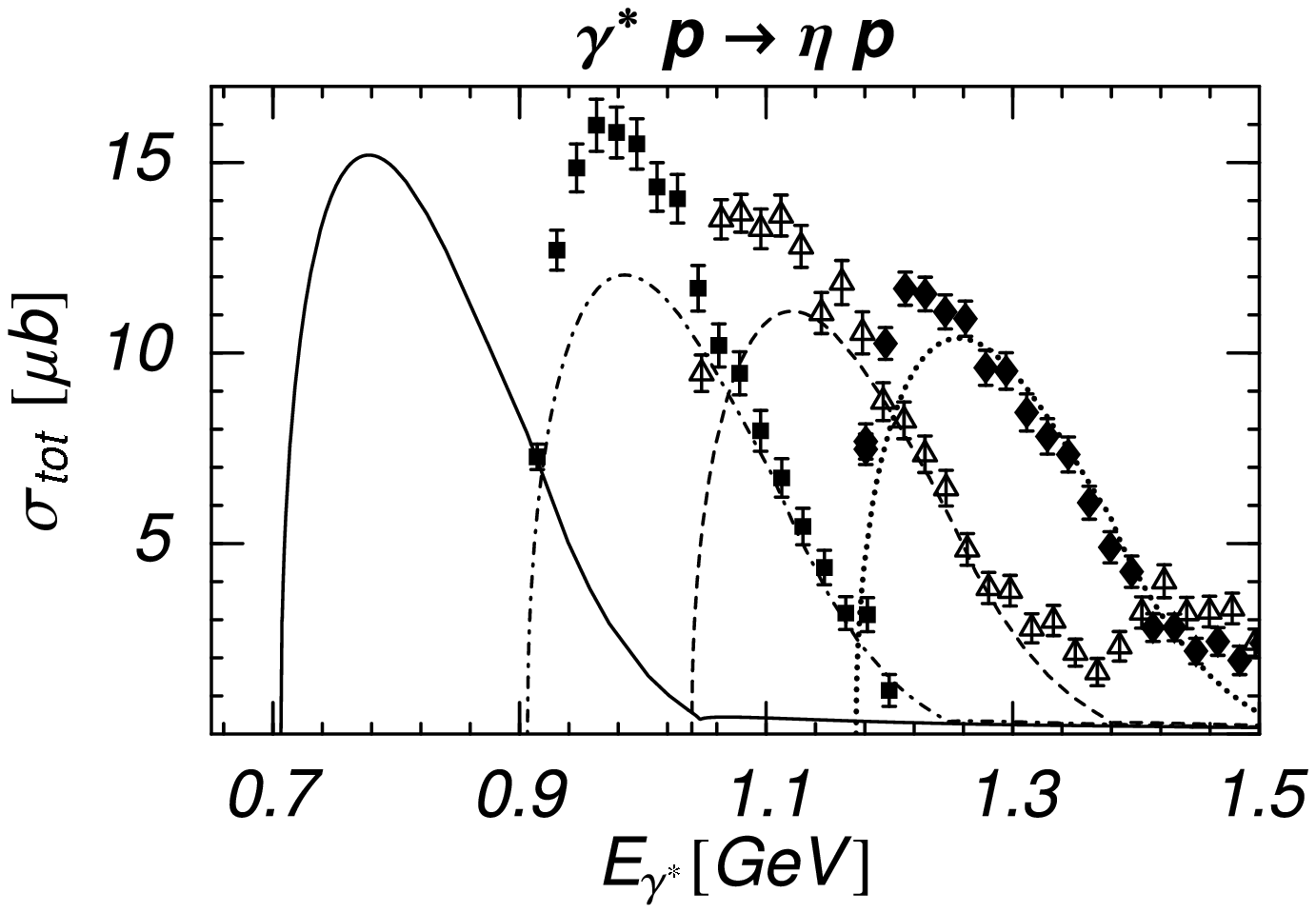,width=7cm}}}
\put(220,20){\makebox(100,120){\epsfig{file=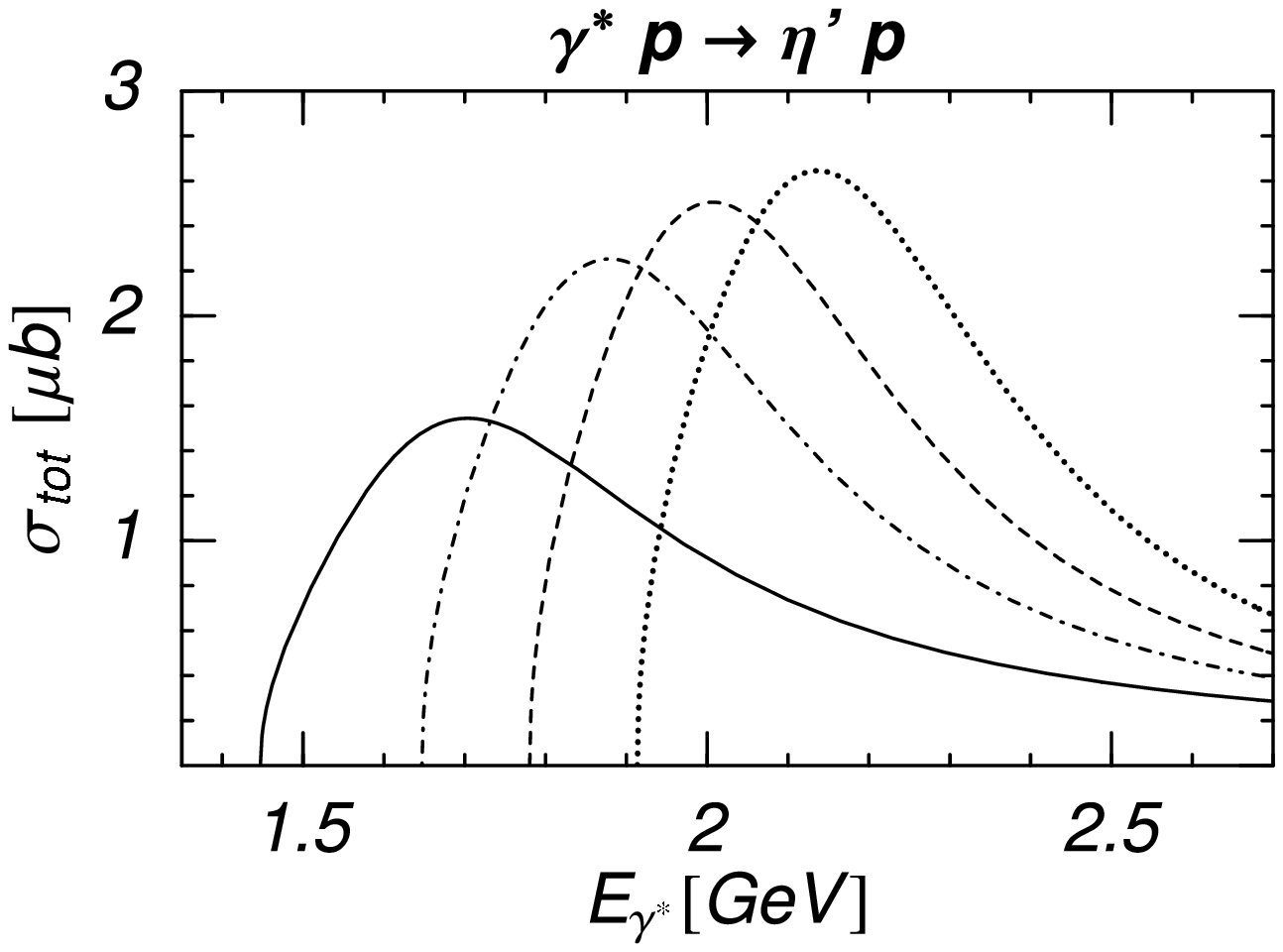,width=7cm}}}
\put(115,115){$a)$}
\put(332,115){$b)$}
\end{picture}
\caption{Cross sections for $\eta$ and $\eta'$ electroproduction on the proton
after the inclusion of electromagnetic form factors for various invariant
momentum transfers $Q^2$. The different lines refer to the
following values of $Q^2$: solid line: photoproduction ($Q^2=0$);
dash-dotted line and squares: $Q^2=0.375$ GeV$^2$; dashed line and
triangles: $Q^2=-0.625$ GeV$^2$; dotted line
and diamonds: $Q^2=0.875$ GeV$^2$. The data for $\eta$ electroproduction 
 with the same $Q^2$ values  
are taken from \cite{thomp}, and only statistical errors are shown.}
\label{fig:elpform}
\end{figure}
In order to model the electromagnetic structure of $B_{0+}$ and $C_{0+}$,
we include  monopole form factors $(1+Q^2/M_\alpha^2)^{-1}$ 
with a mass parameter $M_\alpha$ depending
on the outgoing channel $\alpha$. For $\eta$ and $\eta'$ electroproduction
the channels $|\pi N \rangle$ and $|K \Lambda \rangle$ 
dominate the amplitude.
We choose $M_{|\pi N \rangle} = 0.6$ GeV for the
first channel, so that the form factor is almost identical with the Dirac
form factor of the proton, and $M_{|K \Lambda \rangle} = 2.2$ GeV for the
latter. The remaining three channels play only a minor role
and we set $M_\alpha = 1$ GeV in their case. The results for $\eta$ and
$\eta'$ electroproduction after including these form factors are given in 
Fig.\ \ref{fig:elpform}.
Surprisingly, the cross sections for $\eta$ electroproduction are increased,
while they are reduced in the case of the $\eta'$. This is due to the fact that
we employed different form factors for the participating channels and that
some of these channels may compensate each other in the final state
interactions. To be more precise, for $\eta$ electroproduction the
$|\pi N \rangle$ and $|K \Lambda \rangle$ channels tend to counterbalance
each other, for $\eta'$ electroproduction, on the other hand,
they add up. In contrast to common belief, the inclusion of form factors
can increase the cross section, e.g. in $\eta$ electroproduction, yielding
a $Q^2$ evolution closer to experiment and indicating a hard transition
form factor. The claim that the hard form factor is counterintuitive to
an interpretation of this state as a bound hadronic system \cite{burk}
is not %correct
justified, since within the model the photon couples directly to
one of the ground state octet baryons or  a meson 
and only after this initial 
reaction the produced meson forms a bound state with the baryon.

We do not 
expect our results for electroproduction to reproduce precisely
the experimental data. Nevertheless, the inclusion of simple form factors
for the electroproduction potentials is able to explain
qualitatively the slow
decrease of the $S_{11}(1535)$ photocoupling.
The same form factors also flatten the increase in the $Q^2$ evolution in
$\eta'$ electroproduction but keeping the results still above the cross
section for photoproduction. Again, we cannot regard it as very likely
that the results for the $\eta'$ will be in exact agreement with future
experimental data, in particular due to the uncertainties involved
with the implemented form factors. The qualitative behavior of the $Q^2$
evolution, however, indicates a hard transition form factor also for
$\eta'$ electroproduction. 
Within our model it would be very unlikely to accomodate
both a hard form factor for $\eta$ electroproduction and
faster decreasing cross sections for electroproduction of the $\eta'$.
Upcoming experiments will clarify this issue and 
provide a further test of the model.

\begin{figure}
\centering
%\leavemode
\begin{picture}(330,125)
\put(0,20){\makebox(100,120){\epsfig{file=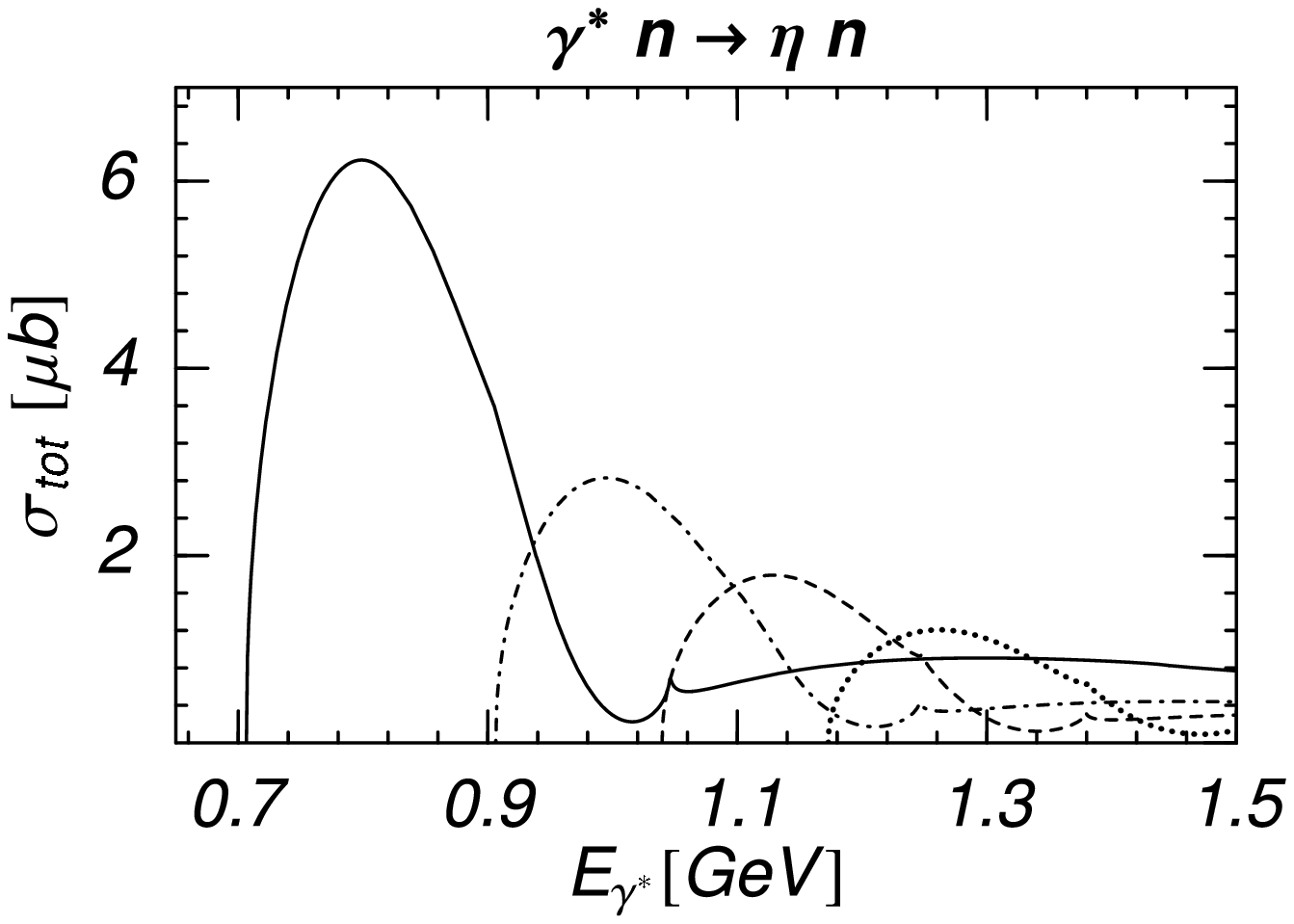,width=7cm}}}
\put(220,20){\makebox(100,120){\epsfig{file=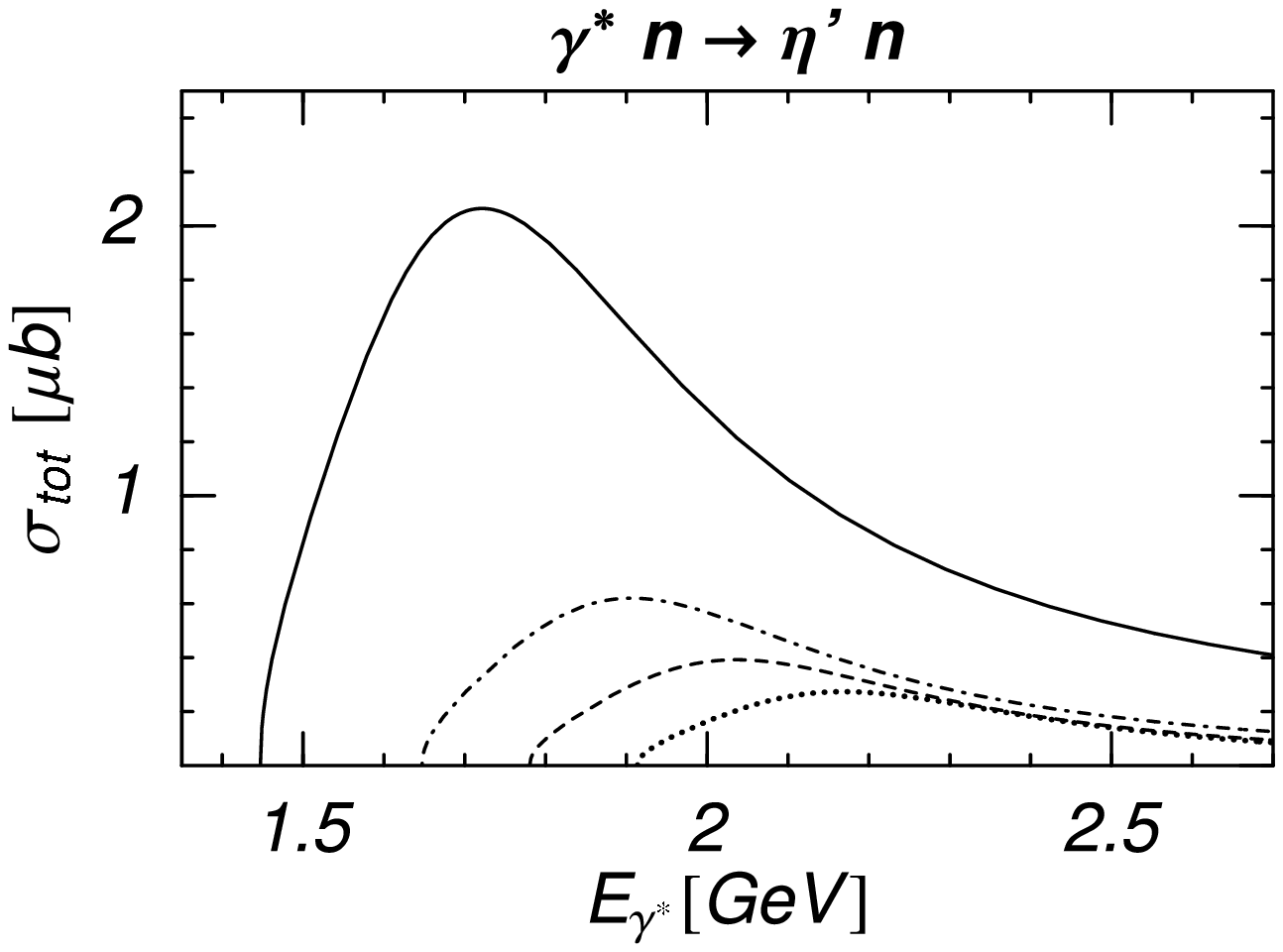,width=7cm}}}
\put(115,115){$a)$}
\put(332,115){$b)$}
\end{picture}
\caption{Cross sections for $\eta$ and $\eta'$ electroproduction on the neutron
for various invariant momentum transfers $Q^2$. The different lines refer
to the following values of $Q^2$: solid line: photoproduction ($Q^2=0$);
dash-dotted line: $Q^2=-0.375$ GeV$^2$; dashed line: $Q^2=-0.625$ GeV$^2$;
dotted line: $Q^2=-0.875$ GeV$^2$.}
\label{fig:elnform}
\end{figure}

Next, we consider $\eta$ and $\eta'$ electroproduction on the neutron, see
Fig.\ \ref{fig:elnform}.
We have employed the same monopole form factors as in the case of the proton.
Photoproduction of the $\eta$ has a cross section of about 6~$\mu b$ at
its peak, being less than half of the cross section on the proton.
This ratio was measured to be approximately 2/3 \cite{etad} which is
clearly above our prediction. This problem was already recognized in
\cite{KWW} and is due to the fact that
the photon cannot couple directly via the charge
to a neutral baryon or meson and thus some of
the electroproduction amplitudes
$B_{0+}$ and $C_{0+}$ vanish, e.g., the $K \Lambda$ channel
to the order we are working. 
To cure this problem, the leading corrections
for the coupling of the photon to a neutral baryon  via its
anomalous magnetic moment
have been taken into account in \cite{KWW} leading to better agreement with
experiment. As these corrections originate from the higher order Lagrangian,
which is beyond the scope of the present investigation, we refrain
from including these terms, but keep in mind that sizable corrections
may occur in electroproduction on the neutron.
The $Q^2$ evolution of the cross section
exhibits a sharp decrease both for $\eta$ and $\eta'$, a behavior which
is remarkably different from the proton case and provides another prediction
to be checked. This is due to the inclusion of form factors, otherwise
the cross sections for $\eta$ and $\eta'$ electroproduction on the neutron,
not shown here for brevity, would grow with $Q^2$.

%%%%%%%%%%%%%%%%%%%%%%%%%%%%%%%%%%%%%%%%%%%%%%%%%%%%%%%%%%%%%%%%%%%%%%%%%%%
\subsection{Effects of the $\eta'$} \label{subsec:effe}

\begin{figure}
\centering
%\leavemode
\begin{picture}(330,125)
\put(0,20){\makebox(100,120){\epsfig{file=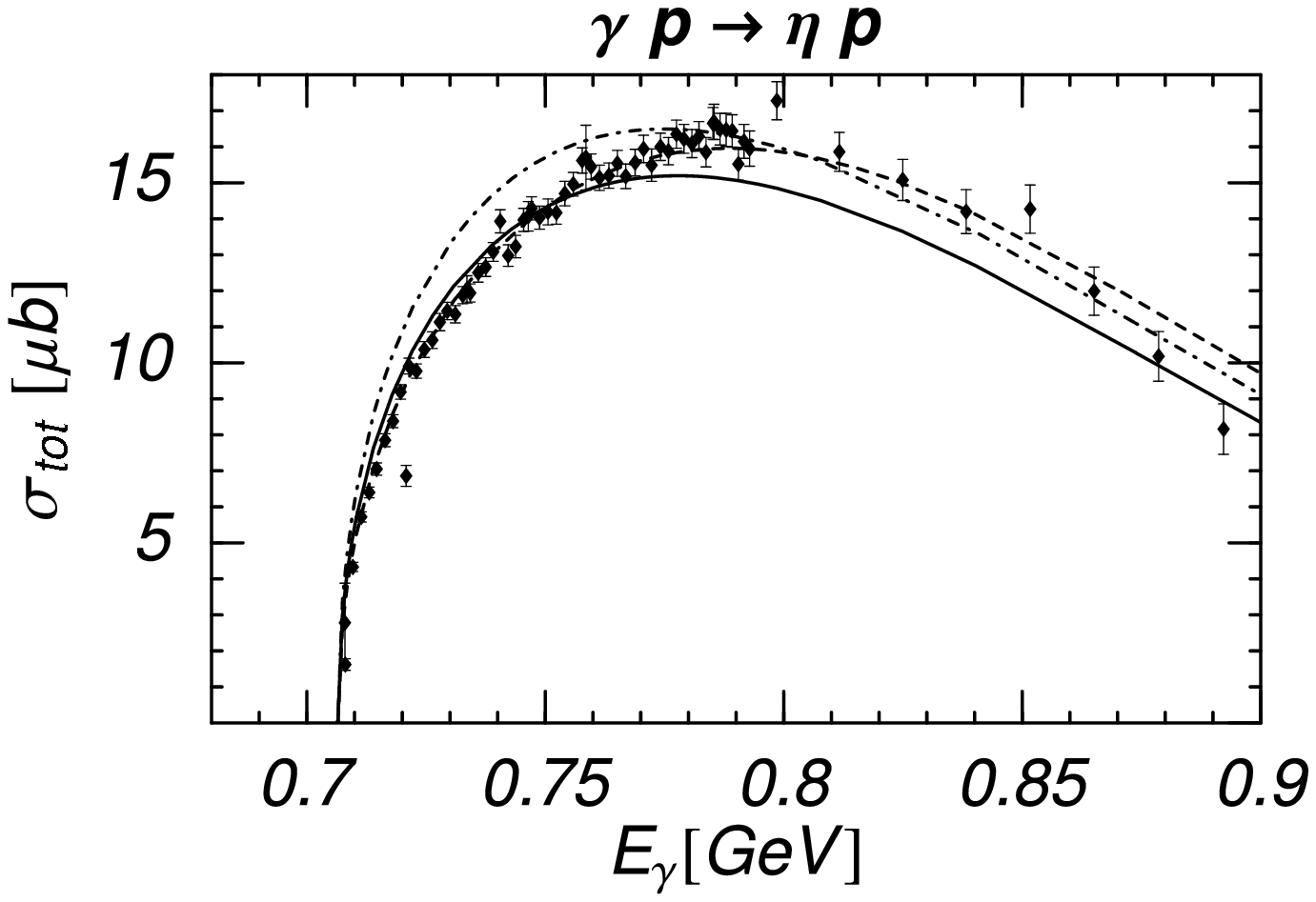,width=7cm}}}
\put(220,20){\makebox(110,120){\epsfig{file=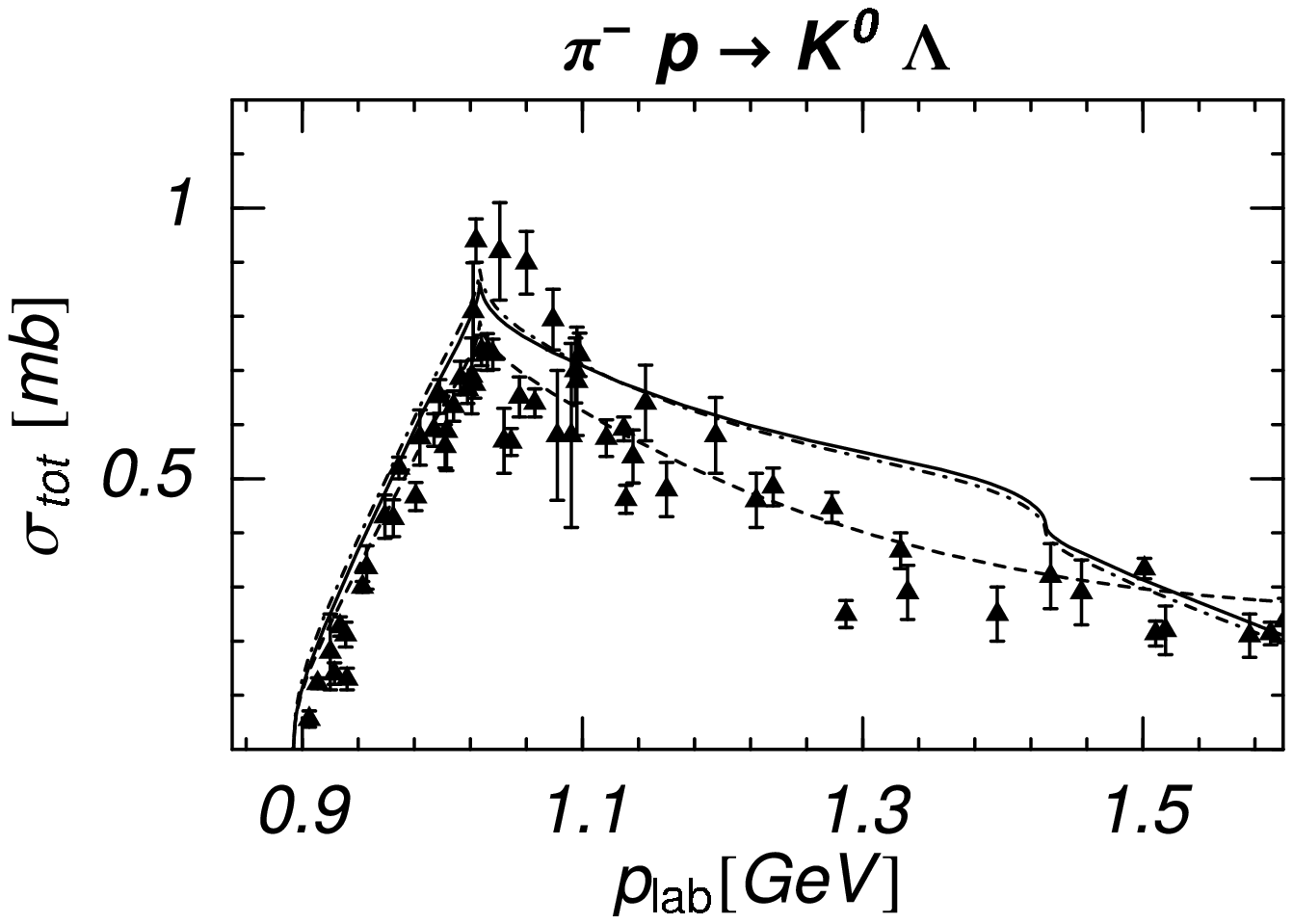,width=7cm}}}
\put(-5,115){$a)$}
\put(225,115){$b)$}
\end{picture}
\caption{Shown are the differences in the cross sections for
$\gamma p \rightarrow  \eta p $ 
and $\pi^- p \rightarrow K^0 \Lambda$ after neglecting $\eta$-$\eta'$ mixing 
and the $| \eta' N \rangle$ channel in the coupled channel formalism.
The solid line is the original result, the dash-dotted line is obtained for
vanishing $\eta$-$\eta'$ mixing, and the dashed line refers to the case without
the $| \eta' N \rangle$ channel.}
\label{fig:effe}
\end{figure}

In this section, we wish to investigate the effects of $\eta$-$\eta'$
mixing and the importance of the $| \eta' N \rangle$ virtual state in our
coupled channel formalism. This is done in a two-step procedure: first,
$\eta$-$\eta'$ mixing is turned off, and secondly, we eliminate the
$| \eta' N \rangle$ channel from the model. In both cases we do not
repeat the fit which would actually compensate most
of the changes. We are particularly interested in the reaction
$\pi^- p \rightarrow K^0 \Lambda$ which exhibits the most prominent
$\eta' N$ cusp of all the channels. Omission of both $\eta$-$\eta'$ mixing 
and the $| \eta' N \rangle$ channel do not lead to substantial
differences in the remaining reactions (where the $\eta'$ is not produced).
In Fig.\ \ref{fig:effe} we have chosen to present in addition to 
$\pi^- p \rightarrow K^0 \Lambda$ the photoproduction process
$\gamma p \rightarrow \eta p$, in order to give a measure for the changes
in the other channels.
For $\pi^- p \rightarrow K^0 \Lambda$ variation in $\eta$-$\eta'$ mixing has
almost no impact (dash-dotted line), like in most other channels in which
the $\eta$ is not
produced as a final particle. Eliminating the $| \eta' N \rangle$ channel 
makes the $\eta' N$ cusp disappear and lowers the cross section, bringing it 
to better agreement with the data (dashed line). It suggests that the
region around the $\eta' N$ cusp is
overemphasized within our model. This feature may change after the inclusion of
$p$-waves, since then a new overall fit to the different reaction channels
will lower the $s$-wave contribution reducing the absolute importance of
the cusp.
For the photoproduction of the $\eta$ on the proton $\eta$-$\eta'$ mixing 
plays a slightly more prominent role (dash-dotted line), as the $\eta$ is
produced in the final state. If the $| \eta' N \rangle$ channel is turned
off, the changes are again quite moderate (dashed line).
Overall we can conclude, that the results for the production of the
Goldstone bosons are not modified substantially after omitting the
$\eta'$ which is in accordance with
intuitive expectation, since the  $| \eta' N \rangle$ channel is much higher
in mass than the other channels. 
We can therefore confirm that it was justified in previous coupled channel
analyses to neglect $\eta$-$\eta'$ mixing and
treat the $\eta$ as a pure octet state, see e.g. \cite{KWW}.

%%%%%%%%%%%%%%%%%%%%%%%%%%%%%%%%%%%%%%%%%%%%%%%%%%%%%%%%%%%%%%%%%%%%%%%%%%%%%%%
\section{Conclusions}

In this work we presented a coupled channel approach to meson-baryon scattering
and electroproduction processes including the $\eta'$ meson 
based on chiral symmetry and unitarity. Since we restrict ourselves to the
threshold region, only $s$-wave contributions
are taken into account which are dominant at the pertinent energies for most 
of the discussed reactions.
The most general chiral effective Lagrangian for the strong interactions 
which includes the $\eta'$ explicitly is
presented up to next-to-leading order and the effective $s$-wave
potentials are derived which are then iterated in a Bethe-Salpeter equation.
Following the work of \cite{KWW} we extended the formalism to include
electroproduction processes where the photon can couple to the hadrons
via their charge.

With only a few chiral parameters --these are, on the one hand, the unknown
coupling constants of the Lagrangian and, on the other hand, the regularization
scales of the loop integrals within the Bethe-Salpeter equation--
we performed a fit to a large amount of data, consisting of meson-baryon
scattering and photoproduction of mesons on the proton.
The $s$-wave approximation yields good agreement with the low-energy data 
in many channels, whereas in the remaining channels $p$-wave contributions
are expected to be dominant. We clearly generate the appearance of the
$S_{11}(1535)$ in the pion- and photo-induced production of the $\eta$,
while the $S_{11}(1650)$ is produced in the $K \Lambda$ channels.
We are also able to produce a resonance-like shape in the
photoproduction of the $\eta'$ which has been assigned to the $S_{11}(1897)$
in a recent experiment at ELSA. 
The overall agreement of our results with data indicates that chiral
dynamics and unitarity
govern processes up to center-of-mass energies of $\sqrt{s} =$2 GeV and that 
the $\eta'$ can be included systematically in a chiral effective
Lagrangian with baryons. 

Having constrained the parameters of the approach, we can give
predictions for further processes such as the pion-induced
production of the $\eta'$ in $\pi^- p \rightarrow \eta' n$ or
$\eta$ and $\eta'$ electroproduction,
which is more sensitive to the structures of the nucleon
due to the longitudinal coupling of the virtual photon to the
nucleon spin and provides a better test for our model when compared with data.
The only available electroproduction data exists so far for the
$\eta$ with invariant momentum transfers $Q^2$ starting at 0.375 GeV$^2$.
It exhibits an unusually  slow 
$Q^2$ evolution which
is in general difficult to understand. Although our results are
larger than that of a simple nucleon dipole form factor, for example,
they exhibit a  faster decrease with $Q^2$ for small values of $Q^2$
which then flattens at higher momentum transfers.
In the case of $\eta'$ electroproduction, the features of the $Q^2$ evolution 
of the cross section are even more striking, since they exhibit a fast
increase instead of the usual decrease.
By choosing different form factors for the initial electroproduction potentials
which take into account the electromagnetic structure of the hadrons
instead of treating them as pointlike particles the cross sections for
$\eta$ electroproduction are {\it increased}, but lowered for the $\eta'$.
This brings the $Q^2$ evolution of $\eta$ electroproduction closer to
experiment and signals a hard transition form factor within the model.
We argue that according to this approach a hard form factor is expected
also for $\eta'$ electroproduction which provides a further test of the model.
In order to make more precise statements, one must include not only $p$-waves,
but also higher order corrections
for the coupling of the photon to a baryon  via its anomalous magnetic moment.
The omission of such contact interactions explains why we fail in reproducing
the photoproduction cross section of the $\eta$ on the neutron.

Finally, we have discussed the impact of $\eta$-$\eta'$ mixing and 
the importance of the $| \eta' N \rangle$ channel within the coupled channel
formalism. Omission of the mixing and the   $| \eta' N \rangle$ channel 
do not lead to substantial changes in those reactions in which the
$\eta'$ is not produced as a final particle. This is in accordance with
intuitive expectation, since the  $| \eta' N \rangle$ channel is much higher
in mass than the other channels.

%%%%%%%%%%%%%%%%%%%%%%%%%%%%%%%%%%%%%%%%%%%%%%%%%%%%%%%%%%%%%%%%%%%%%%%%%%%%%%%
\section*{Acknowledgements}

We are grateful to N. Kaiser who shared his insight in  the coupled channel 
formalism with us. Useful discussions with T. Hemmert and W. Weise are
gratefully acknowledged. We would also like to thank J. Mueller for providing 
us with the data for $\eta$ electroproduction from CLAS at JLab.
This work has been supported in part by the Deutsche Forschungsgemeinschaft.

%%%%%%%%%%%%%%%%%%%%%%%%%%%%%%%%%%%%%%%%%%%%%%%%%%%%%%%%%%%%%%%%%%%%%%%%%%%%%%
\appendix 

%%%%%%%%%%%%%%%%%%%%%%%%%%%%%%%%%%%%%%%%%%%%%%%%%%%%%%%%%%%%%%%%%%%%%%%%%%%%%%%
\section{} \label{app:pot}
In this appendix, we list the coefficients $C_{\alpha}^{\beta}(s)$ 
from Eq. (\ref{ampl}) in the isospin basis.
There are five channels with total isospin $I= 1/2$ which
are $|\pi N \rangle^{(1/2)}$, $|\eta N \rangle^{(1/2)}$,
$|K \Lambda \rangle^{(1/2)}$,
$|K \Sigma \rangle^{(1/2)}$, $|\eta' N \rangle^{(1/2)}$
(labelled with indices 1, 2, 3, 4, 5, respectively) and
two channels with $I=3/2$,
$|\pi N \rangle^{(3/2)}$, $|K \Sigma \rangle^{(3/2)}$
(labeled by indices 6 and 7). 
The physical states can be expressed in terms of the isospin states,
\begin{eqnarray}
|\pi^+ n \rangle &=& \frac{1}{\sqrt{3}} \Big( \sqrt{2} |\pi N \rangle^{(1/2)}
                      + |\pi N \rangle^{(3/2)} \Big)       \no \\
|\pi^0 p \rangle &=& \frac{1}{\sqrt{3}} \Big( |\pi N \rangle^{(1/2)}
                      - \sqrt{2} |\pi N \rangle^{(3/2)} \Big)       \no \\
|K^+ \Sigma^0 \rangle &=& \frac{1}{\sqrt{3}} \Big( |K \Sigma \rangle^{(1/2)}
                      - \sqrt{2} |K \Sigma \rangle^{(3/2)}\Big)       \no \\
|K^+ \Sigma^- \rangle &=& \frac{1}{\sqrt{3}} \Big( \sqrt{2}|K \Sigma 
\rangle^{(1/2)}
                      -  |K \Sigma \rangle^{(3/2)}\Big)       \no \\
|K^0 \Sigma^+ \rangle &=& \frac{1}{\sqrt{3}} \Big( \sqrt{2}|K \Sigma 
\rangle^{(1/2)}
                      +  |K \Sigma \rangle^{(3/2)}\Big)       \no \\
|K^0 \Sigma^0 \rangle &=& \frac{1}{\sqrt{3}} \Big( -|K \Sigma \rangle^{(1/2)}
                      - \sqrt{2} |K \Sigma \rangle^{(3/2)}\Big)   ,
\end{eqnarray}                      
where we have used the convention in which for the following
states there is a sign difference between the physical and isospin basis:
$|\pi^+\rangle = - |1,1\rangle$, $|K^+\rangle = - |1/2,1/2 \rangle$,
$|K^0\rangle = - |1/2, -1/2 \rangle$, $|\bar{K}^0\rangle = - |1/2,1/2\rangle$,
$|\Sigma^+\rangle = - |1,1\rangle$, $|p\rangle = - |1/2,1/2 \rangle$,
$|n\rangle = - |1/2, -1/2 \rangle$, $|\Xi^0\rangle = - |1/2,1,2\rangle$.

The octet field $\eta_8$ and its singlet counterpart $\eta_0$ as they appear
in the Lagrangian are related to the physical mass eigenstates $\eta$ and
$\eta'$ in the one-mixing-angle scheme as follows
\begin{eqnarray} \label{App:mix}
\eta_8 &=& \eta \, \cos \vartheta + \eta' \, \sin \vartheta \no \\
\eta_0 &=& -\eta \,\sin \vartheta + \eta' \, \cos \vartheta ,
\end{eqnarray}
where we have used $\vartheta = -10^{\circ}$. This is the value
which one obtains from the Gell-Mann--Okubo mass relation for the
pseudoscalar mesons at lowest order. As discussed in the main text,
small variations in this value lead only to moderate changes which
can even be partially compensated by repeating the fit.
The implementation of the more general two-mixing-angle scheme,
see e.g. \cite{BB}, does not change
any of our conclusions and will therefore not be considered here.

The coefficients $C_i^j$ read
\begin{eqnarray}
C_1^1 &=&  \sqrt{s} - M_N - \frac{3}{4}(D+F)^2 \,
\frac{(\sqrt{s} -M_N)^2
}{\sqrt{s} + M_N}  \no \\
&&           - 2 ( 2 b_0 + b_D + b_F) m_\pi^2  
            + 2 (d_1 + d_2 + 2d_4) P_1^1 \no \\
C_1^2 &=& -\frac{1}{4} (D+F) \Big( \sqrt{2} (2D + 3D_s) \sin \vartheta 
+ (D-3F) \cos \vartheta \Big) \frac{(\sqrt{s} -M_N)^2}{\sqrt{s} 
+ M_N}  \no \\
&&           + 2 \Big( [b_D + b_F] \cos \vartheta - \big[ \sqrt{2} (b_D 
+ b_F) - \sqrt{3} (c_D + c_F) \big] \sin \vartheta \Big) m_\pi^2 \no \\
&&          - \Big( 2 \big[ d_1 + 3d_2\big] \cos \vartheta
-\sqrt{2} \big[ 2 d_1 + 3d_5 + 3 d_6\big] 
\sin \vartheta  \Big) P_1^2\no \\
C_1^3 &=&  - \frac{3}{8}  [ M_N + M_\Lambda - 2 \sqrt{s}] 
         - \frac{1}{4} (D+F) (D+3F)  \frac{(\sqrt{s} - M_N)(\sqrt{s} 
- M_\Lambda)}{\sqrt{s} + M_N}  \no \\
&&           - \frac{1}{2} (b_D +3 b_F) ( m_K^2 + m_\pi^2)
            +3(d_1 + d_2 ) P_1^3  \no \\ 
C_1^4 &=& - \frac{1}{8}  [ M_N + M_\Sigma - 2 \sqrt{s} ]
         - \frac{3}{4} (D^2-F^2)  \frac{(\sqrt{s} - M_N)(\sqrt{s} -
 M_\Sigma)}{\sqrt{s} + M_N}   \no \\
&&              +  \frac{1}{2} (b_D - b_F) ( m_K^2 + m_\pi^2) 
        + (d_1 - 7 d_2 + 2d_3) P_1^4 \no \\
C_1^5 &=& \frac{1}{4}  (D+F) \Big( \sqrt{2} (2D + 3D_s) \cos \vartheta -
 (D-3F) \sin \vartheta \Big)  \frac{(\sqrt{s} -M_N)^2}{\sqrt{s}
 + M_N} \no \\
&&           + 2 \Big( [b_D + b_F] \sin \vartheta + \big[ 
\sqrt{2} (b_D + b_F) - \sqrt{3} (c_D + c_F) \big] \cos \vartheta
 \Big) m_\pi^2 \no \\
&&          - \Big( \sqrt{2} \big[ 2 d_1 + 3d_5 + 3 d_6\big]
 \cos \vartheta + 2 \big[ d_1 + 3d_2\big] \sin \vartheta
 \Big) P_1^5\no \\
C_2^2 &=&  -\frac{1}{12} \Big( \sqrt{2} (2D+3D_s) \sin \vartheta 
+ (D-3F) \cos \vartheta \Big)^2  \frac{(\sqrt{s} -M_N)^2}{\sqrt{s}
 + M_N}    - 4 M_N u_1 \sin^2 \vartheta  \no \\
&&             + \frac{8}{3} \Big( - \big[ b_0 + b_D -b_F \big]
 \big( \sqrt{2} \cos \vartheta + \sin \vartheta \big)^2 
           +  \sqrt{6} \big[ c_0 +c_D - c_F \big] \sin \vartheta 
\big( \sqrt{2} \cos \vartheta + \sin \vartheta \big) \Big)
 m_K^2 \no \\   
&&           + \frac{2}{3} \Big( \big[ 2 b_0 + 3b_D -5b_F \big] 
\cos^2 \vartheta - 2 \big[ \sqrt{2} ( -4b_0 -3b_D + b_F) 
             +  \sqrt{3} ( 4c_0 +3c_D - c_F) \big] \cos \vartheta 
\sin \vartheta \no \\
&&             - \frac{4}{3} \big[ b_0 + 2 b_F - \sqrt{6} ( c_0 + 2 c_F) \big] 
\sin^2 \vartheta   \Big) m_\pi^2  \no \\
&&          + 2 \Big( 2 \big[ d_4 + 2d_5 + 3 d_7\big] \sin^2 \vartheta
 - \sqrt{2} \big[ 2 d_1 -d_5 + 3 d_6\big] \sin \vartheta \cos \vartheta 
             + \big[ -d_1 + 3d_2 + 2 d_4\big] \cos^2 \vartheta \Big)
 P_2^2\no %\\
\end{eqnarray}
\begin{eqnarray}
C_2^3 &=&  \frac{3}{8}  [ M_N + M_\Lambda - 2 \sqrt{s}]\cos \vartheta \no \\
&&           - \frac{1}{12} (D+3F) \Big( \sqrt{2} (2D + 3D_s)
 \sin \vartheta + (D-3F) \cos \vartheta \Big) 
 \frac{(\sqrt{s} - M_N)(\sqrt{s} - M_\Lambda)}{\sqrt{s} + M_N}   \no \\
&&           + \frac{1}{6}  [b_D +3 b_F] ( 3 m_\pi^2 - 5 m_K^2 )  
\cos \vartheta - \frac{2}{3} \big[  \sqrt{2} (b_D +3 b_F) -
 \sqrt{3} (c_D +3 c_F) \big] m_K^2 \sin \vartheta  \no \\
&&          + \Big( \sqrt{2} \big[ 2 d_1 +d_5 + 3 d_6\big] 
\sin \vartheta +  \big[ d_1 - 3d_2 + 2 d_3\big] 
\cos \vartheta \Big) P_2^3\no \\
C_2^4 &=&  \frac{3}{8} [2\sqrt{s} -M_N-M_\Sigma] \cos \vartheta \no \\
&&           - \frac{1}{4} (D-F) \Big( \sqrt{2} (2D + 3D_s) 
\sin \vartheta + (D-3F) \cos \vartheta \Big)  
\frac{(\sqrt{s} - M_N)(\sqrt{s} - M_\Sigma)}{\sqrt{s} + M_N}   \no \\
&&           + \frac{1}{2}  [b_D - b_F] ( 3 m_\pi^2 - 5 m_K^2 )  
\cos \vartheta - 2 \big[  \sqrt{2} (b_D - b_F) - \sqrt{3} 
(c_D - c_F) \big] m_K^2 \sin \vartheta  \no \\
&&          - \Big( \sqrt{2} \big[ 2 d_1 -3d_5 + 3 d_6\big] 
\sin \vartheta +  \big[ d_1 - 3d_2\big] \cos \vartheta \Big) P_2^4\no \\
C_2^5 &=&  \frac{1}{12}  \big(  \sqrt{2} (2D+3D_s) \sin \vartheta 
+ (D-3F) \cos \vartheta \big)
           \big( \sqrt{2} (2D+3D_s) \cos \vartheta - (D-3F) 
\sin \vartheta \big)  \no \\
&&          \times  \frac{(\sqrt{s} -M_N)^2}{\sqrt{s} + M_N} 
      + 4 M_N u_1 \cos \vartheta \sin \vartheta \no \\
&&   + \frac{1}{3} \bigg[
\Big(8\sqrt{2}(b_0 +b_D - b_F) -8 \sqrt{3} (c_0 +c_D - c_F) \Big) m_K^2\no \\
&&  -\Big(2 \sqrt{2}(4b_0 +3b_D - b_F) -2 \sqrt{3} ( 4c_0 +3c_D - c_F) \Big)
 m_\pi^2  \no \\
&&+3 \sqrt{2} (2 d_1 -d_5 + 3 d_6) P_2^5\bigg]\cos (2\vartheta) \no \\
&&-\frac{1}{3} \bigg[
4\Big(b_0 +b_D - b_F + \sqrt{6}(c_0 +c_D - c_F)\Big) m_K^2 \no \\
&&+\Big(-4 b_0 - 3 b_D + b_F + 2 \sqrt{6} (c_0 +2 c_F)\Big) m_\pi^2 \no \\
&&+ 3(d_1 -3 d_2 +4 d_5 + 6 d_7)P_2^5 \bigg] \sin (2 \vartheta)
\end{eqnarray}

\begin{eqnarray} 
C_3^3 &=&     - \frac{1}{12} (D+3F)^2  \frac{(\sqrt{s} -
M_\Lambda)^2}{\sqrt{s} + M_N} \no \\
&&        - \frac{2}{3} ( 6 b_0 + 5 b_D) m_K^2  +
2 (3 d_2 + 2 d_4) P_3^3   \no \\
C_3^4 &=&     - \frac{1}{4} (D-F) (D-3F)  \frac{(\sqrt{s} -
M_\Sigma)(\sqrt{s} - M_\Lambda)}{\sqrt{s} + M_N}  \no \\
&&     + 2b_D m_K^2 - 6 d_2 P_3^4\no \\
C_3^5 &=&  \frac{3}{8} [ M_N + M_\Lambda - 2 \sqrt{s}]  \sin \vartheta\no \\
&&           + \frac{1}{12} (D+3F) \Big( \sqrt{2} (2D + 3D_s)
\cos \vartheta - (D-3F) \sin \vartheta \Big) 
\frac{(\sqrt{s} - M_N)(\sqrt{s} - M_\Lambda)}{\sqrt{s} + M_N}   \no \\
&&           + \frac{1}{6}  [b_D +3 b_F] ( 3 m_\pi^2 - 5 m_K^2 )
 \sin \vartheta + \frac{2}{3} \big[  \sqrt{2} (b_D +3 b_F) -
\sqrt{3} (c_D +3 c_F) \big] m_K^2 \cos \vartheta  \no \\
&&          - \Big( \sqrt{2} \big[ 2 d_1 +d_5 + 3 d_6\big]
\cos \vartheta -  \big[ d_1 - 3d_2 + 2 d_3\big]
\sin \vartheta \Big) P_3^5\no \\
C_4^4 &=&  \sqrt{s} - M_\Sigma 
           - \frac{3}{4} (D-F)^2  \frac{(\sqrt{s} -
M_\Sigma)^2}{\sqrt{s} + M_N}   \no \\
&&          - 2 ( 2 b_0 + b_D - 2 b_F) m_K^2   
      + 2( - 2d_1 + d_2 + 2 d_4) P_4^4  \no \\                      \no \\
C_4^5 &=&  \frac{3}{8}[2\sqrt{s} -M_N-M_\Sigma] \sin \vartheta\no \\
&&           + \frac{1}{4} (D-F) \Big( \sqrt{2} (2D + 3D_s)
\cos \vartheta - (D-3F) \sin \vartheta \Big) 
\frac{(\sqrt{s} - M_N)(\sqrt{s} - M_\Sigma)}{\sqrt{s} + M_N}   \no \\
&&           + \frac{1}{2}  [b_D - b_F] ( 3 m_\pi^2 - 5 m_K^2 ) 
\sin \vartheta + 2 \big[  \sqrt{2} (b_D - b_F) - \sqrt{3}
(c_D - c_F) \big] m_K^2 \cos \vartheta  \no \\
&&          + \Big( \sqrt{2} \big[ 2 d_1 -3d_5 + 3 d_6\big]
\cos \vartheta -  \big[ d_1 - 3d_2\big] \sin \vartheta
\Big) P_4^5\no \\
C_5^5 &=&  -\frac{1}{12}   \Big( \sqrt{2} (2D+3D_s) \cos \vartheta
- (D-3F) \sin \vartheta \Big)^2 
       \frac{(\sqrt{s} -M_N)^2}{\sqrt{s} + M_N}
      - 4 M_N u_1 \cos^2 \vartheta  \no \\ 
&&             + \frac{8}{3} \Big( - \big[ b_0 + b_D -b_F \big]
\big( \sqrt{2} \sin \vartheta - \cos \vartheta \big)^2 
           -  \sqrt{6} \big[ c_0 +c_D - c_F \big] \cos \vartheta
\big( \sqrt{2} \sin \vartheta - \cos \vartheta \big) \Big)
m_K^2 \no \\   
&&           + \frac{2}{3} \Big( \big[ 2 b_0 + 3b_D -5b_F \big]
\sin^2 \vartheta + 2 \big[ \sqrt{2} ( -4b_0 -3b_D + b_F) 
             +  \sqrt{3} ( 4c_0 +3c_D - c_F) \big] \sin \vartheta
\cos \vartheta \no \\
&&             - 2 \big[ b_0 + 2 b_F - \sqrt{6} ( c_0 + 2 c_F) \big]
\cos^2 \vartheta   \Big) m_\pi^2  \no \\
&&          + 2 \Big( 2 \big[ d_4 + 2d_5 + 3 d_7\big]
\cos^2 \vartheta + \sqrt{2} \big[ 2 d_1 -d_5 + 3 d_6\big]
\cos \vartheta \sin \vartheta 
             + \big[ -d_1 + 3d_2 + 2 d_4\big] \sin^2 \vartheta
\Big) P_5^5\no %\\
\end{eqnarray}
\begin{eqnarray}
C_6^6 &=&  - \frac{1}{2} [\sqrt{s} - M_N] 
- 2 ( 2 b_0 + b_D + b_F) m_\pi^2  
                      + 2 (d_1 + d_2 + 2d_4) P_6^6\no \\
C_6^7 &=&  \frac{1}{4}  [ M_N + M_\Sigma - 2 \sqrt{s} ]    
    - (b_D - b_F) ( m_K^2 + m_\pi^2) 
                     + 2 (-d_1 + d_2 + d_3) P_6^7\no \\
C_7^7 &=&  - \frac{1}{2} [ \sqrt{s} - M_\Sigma] 
          - 2 ( 2 b_0 + b_D +b_F) m_K^2 
                      + 2 (d_1 + d_2 + 2d_4) P_7^7\no %\\
\end{eqnarray}
For an isospin state labeled $\alpha$ ($\beta$) which consists of a
baryon $a$ ($b$) and a meson $i$ ($j$) we have used the abbreviation
\begin{equation}
P_\alpha^\beta = E_i E_j + \frac{|\mbox{\boldmath$p$}|^2
|\mbox{\boldmath$p'$}|^2}{ 3 N_a^2 N_b^2} .
\end{equation}
The energies of the mesons are given by $E_{i (j)}$
and $\mbox{\boldmath$p$}$ ($\mbox{\boldmath$p'$}$) is the
three-momentum of the
incoming (outgoing) baryon in the center-of-mass frame. 

%%%%%%%%%%%%%%%%%%%%%%%%%%%%%%%%%%%%%%%%%%%%%%%%%%%%%%%%%%%%%%%%%%%%%%%%%%%%%%


\begin{thebibliography}{99}



\bibitem{KSW1} 
N. Kaiser, P. B. Siegel, W. Weise, Nucl. Phys. {\bf A594} (1995) 325

\bibitem{KSW2} 
N. Kaiser, P. B. Siegel, W. Weise, Phys. Lett. {\bf B362} (1995) 23

\bibitem{OO} J. A. Oller, E. Oset, Nucl. Phys. {\bf A620} (1997) 438;\\
             J. A. Oller, E. Oset, J. R. Pel{\'a}ez, Phys. Rev.
{\bf D59} (1999) 074001

\bibitem{OR} E. Oset, A. Ramos, Nucl. Phys. {\bf A635} (1998) 99


\bibitem{OM} J. A. Oller, U.-G. Mei{\ss}ner, Phys. Lett. {\bf B500} (2001) 263

\bibitem{Inoue} T. Inoue, E. Oset, M. J. Vicente Vacas, Phys. Rev.
{\bf C65} (2002) 035204

\bibitem{mami}
B. Krusche et al., Phys. Rev. Lett. {\bf 74} (1995) 3736


\bibitem{elsa}
B. Schoch, Prog. Part. Nucl. Phys. {\bf 35} (1995) 43


\bibitem{CP} C. E. Carlson, J. L. Poor,  Phys. Rev. {\bf D38} (1988) 2758

\bibitem{LDW} D. B. Leinweber, T. Draper, R. M. Woloshyn, Phys. Rev. {\bf D46} 
(1992) 3067

\bibitem{eldel} V. V. Frolov et al.,  Phys. Rev. Lett. {\bf 82}, (1999) 45

\bibitem{els11} C. S. Armstrong et al.,  Phys. Rev. {\bf D60}, (1999) 052004

\bibitem{thomp} R. Thompson et al.,  Phys. Rev. Lett. {\bf 86}, (2001) 1702


\bibitem{BeM} M. Benmerrouche, N. C. Mukhopadhyay,  
              Phys. Rev. Lett. {\bf 67} (1991) 1070; \\
              M. Benmerrouche, N. C. Mukhopadhyay, J. F. Zhang, 
              Phys. Rev. {\bf D51} (1995) 3237

\bibitem{Be} C. Bennhold, H. Tanabe, Nucl. Phys. {\bf A530} (1991) 625

\bibitem{Ti} L. Tiator, C. Bennhold, S.S. Kamalov, 
             Nucl. Phys. {\bf A580} (1994) 455

\bibitem{KWW}  N. Kaiser, T. Waas, W. Weise, Nucl. Phys. {\bf A612} (1997) 
               297; \\
               J. Caro Ramon, N. Kaiser, S. Wetzel, W. Weise, 
               Nucl. Phys. {\bf A672} (2000) 249


\bibitem{W} S. Weinberg, Phys. Rev. {\bf D11} (1975) 3583

\bibitem{leut} H. Leutwyler, Phys. Lett. {\bf B374} (1996) 163

\bibitem{H-S} P.~Herrera-Sikl\'ody, J.~I.~Latorre, P.~Pascual and J.~Taron,
                Nucl. Phys. {\bf B497} (1997) 345

\bibitem{B} B. Borasoy, Phys. Rev. {\bf D61} (2000)  014011


\bibitem{ZMB} J. F. Zhang, N. C. Mukhopadhyay, M. Benmerrouche, 
                Phys. Rev. {\bf C52} (1995) 1134

\bibitem{Li}  Z. Li, J. Phys. {\bf G23} (1997) 1127


\bibitem{BWW} S. D. Bass, S. Wetzel, and W. Weise, Nucl. Phys.
{\bf A686} (2001) 429 

\bibitem{Pl}  R. Pl{\"o}tzke et al., Phys. Lett. {\bf B444} (1998) 555


\bibitem{N} B. Norum, private communication;\\ E. Pasyuk, private communication


\bibitem{KL} R. Kaiser, H. Leutwyler, Eur. Phys. J. {\bf C17} (2000) 623 


\bibitem{BB} N. Beisert, B. Borasoy, Eur. Phys. J. {\bf A11} (2001) 329 


\bibitem{CR} F. E. Close, R. G. Roberts, Phys. Lett. {\bf B316} (1993) 165;\\
             B. Borasoy, Phys. Rev. {\bf D 59} (1999) 054021


\bibitem{GMO} M. Gell-Mann, Phys. Rev. {\bf 125} (1962) 1067;\\
              S. Okubo, Prog. Theor. Phys.  {\bf 27} (1962) 949


\bibitem{GLS} J. Gasser, H. Leutwyler, M.E. Sainio, Phys.Lett.
{\bf B253} (1991) 252 


\bibitem{PSWA}  M.M. Pavan, I.I. Strakovsky, R.L. Workman, R.A. Arndt,  
"{\it The pion nucleon sigma term is definitely large: results
from a G.W.U. analysis
of pion nucleon scattering data.}", hep-ph/0111066 


\bibitem{BM} B. Borasoy, U.-G. Mei{\ss}ner, Ann. Phys. (NY)
{\bf 254} (1997) 192 


\bibitem{BB2} N. Beisert, B. Borasoy, Nucl. Phys. {\bf A705} (2002) 433 


\bibitem{FK} T. Feldmann, P. Kroll, Eur. Phys. J. {\bf C5} (1998) 327 


\bibitem{BL} T. Becher, H. Leutwyler, Eur. Phys. J. {\bf C9} (1999) 643 


\bibitem{NA} J. Nieves and E. Ruiz Arriola, Nucl. Phys.  {\bf A679} (2000) 57

\bibitem{Belectro} B. Borasoy, Phys.\ Rev.\ {\bf D63}, (2001) 094015

\bibitem{bald}  A. Baldini et al., in: Landolt-B\"ornstein (Ed.), Vol. 
12a, Springer, Berlin, 1988

\bibitem{tran} M. Q. Tran et al. (SAPHIR Collaboration), Phys. Lett. 
{\bf B445} (1998) 20

\bibitem{goers} S. Goers et al. (SAPHIR Collaboration), Phys. Lett. 
{\bf B464} (1999) 331

\bibitem{gian} M. M. Giannini, E. Santopinto, A. Vasallo, Nucl. Phys. 
{\bf A699} (2002) 308

\bibitem{burk} V. D. Burkert, Talk presented at Baryons 2002, Newport News,
     USA,            to appear in the proceedings; hep-ph/0207149

\bibitem{etad}  P. Hoffmann-Rothe et al., Phys. Rev. Lett. {\bf 78} (1997) 4697


\end{thebibliography}
\end{document}